\documentclass[preprint,11pt]{elsarticle}

\usepackage{float}
\usepackage{graphicx}
\usepackage{latexsym}
\usepackage{amsmath}
\usepackage{mathrsfs}
\usepackage{color}
\usepackage{tabls}
\usepackage{graphicx}
\usepackage{epsfig}
\usepackage{subfigure}
\usepackage{rotating}
\usepackage{latexsym}
\usepackage{amsmath}
\usepackage[mathcal]{eucal}
\usepackage{amssymb}
\usepackage{colortbl}
\usepackage{times}
\usepackage{longtable}
\usepackage{calc}
\usepackage{algorithmic}
\usepackage{algorithm}
\usepackage{psfrag}
\usepackage{tabularx}
\usepackage{tikz}
\usetikzlibrary{arrows.meta,positioning,shapes.geometric}
\usetikzlibrary{arrows.meta,positioning,matrix}
\usetikzlibrary{arrows.meta,positioning}

\usepackage{bbm}
\usepackage{stmaryrd}
\usepackage{graphicx}
\usepackage{natbib}
\usepackage{caption}

\usepackage{setspace}
\usepackage{fullpage}

\newcommand{\be}{\begin{eqnarray}}
\newcommand{\ee}{\end{eqnarray}}
\newcommand{\ben}{\begin{eqnarray*}}
\newcommand{\een}{\end{eqnarray*}}

\def\det{\mathrm {det}}

\def\det{\mathrm {det}}

\journal{International Journal of Engineering Science}
\begin{document}

\begin{frontmatter}

\title{Constitutive modelling of magneto-active polymers at finite strains: A survey}

\author[add1]{Abhishek Ghosh}
\ead{abhishek.ghosh@swansea.ac.uk}
\author[add2]{Chennakesava Kadapa}
\ead{c.kadapa@napier.ac.uk}
\author[add1]{Mokarram Hossain \corref{cor1}}
\ead{mokarram.hossain@swansea.ac.uk}

\address[add1]{ Zienkiewicz Institute for Modelling, Data and AI, Swansea University, Bay Campus, SA1 8EN, Swansea, UK}
\address[add2]{School of Computing, Engineering and the Built Environment, Edinburgh Napier University, Edinburgh, EH10 5DT, UK
}
\cortext[cor1]{Corresponding author. Tel.: +44 074-82959957; Fax: +49 09131-8528503.}

\begin{abstract}

\noindent Magneto-active polymers (MAPs) are field-responsive soft composites whose mechanical behaviour can be actively modified by external magnetic fields. Their ability to exhibit field-induced stiffening, magnetostriction, anisotropy and rate-dependent response makes them attractive for sensors, actuators, adaptive structures, vibration-control devices and soft robotic systems. This article presents a structured survey of constitutive modelling approaches for MAPs at finite strains. The review traces the development from early semi-empirical descriptions to thermodynamically consistent nonlinear continuum theories, with particular attention to isotropic and anisotropic magnetoelastic constitutive models, invariant-based and spectral representations, variational and polyconvex frameworks, microstructurally motivated models, dispersed-chain descriptions and magneto-viscoelastic theories. The principal constitutive variables, energy functions, coupling mechanisms and physical assumptions underlying the main models are discussed. The survey shows that constitutive modelling of MAPs has developed into a broad family of nonlinear, anisotropic and dissipative frameworks capable of describing versatile behaviours observed experimentally. Important challenges remain in resolving thermo-magneto-mechanical coupling and microstructure-sensitive effects, identification of parameters, validation of models as well as robust and efficient implementation on computers.

\end{abstract}


\end{frontmatter}

\section{Introduction}

\noindent Magneto-active polymers (MAPs) are field-responsive soft composites whose mechanical behaviour can be actively controlled by externally applied magnetic fields. They are commonly formed by embedding magnetisable micro- or nano-particles within a compliant polymeric matrix. Under magnetic loading, interactions between the particles and the polymer network modify the macroscopic response, leading to field-dependent stiffness, magnetostriction, magnetic actuation, anisotropy and time-dependent behaviour \cite{jolly1,ginder1,dorfmann3,dorfmann4}. Owing to their softness, reversibility and remote actuation capability, MAPs have attracted considerable interest in adaptive structures, vibration-control devices, soft actuators, sensors, biomedical systems and soft robotic technologies \cite{alba03,deng1,fars04,ginder2,ginder3,kashima1,yalcintas1,zhu1,bose2,bose3,bastola2,zhao2019mechanics}.

\noindent A useful broad distinction can be made between soft-magnetic and hard-magnetic soft materials depending on the magnetisation mechanisms and responsive behaviour. Soft-magnetic soft materials contain low-coercivity particles, such as iron or iron oxides, whose magnetisation is mainly induced by the applied magnetic field and changes appreciably with the field. Their response is therefore governed by field-induced particle interactions, magnetisation processes and the evolving magnetic state of the composite \cite{jolly1,ginder1,danas1,dorfmann3,dorfmann4,kankanala1}. By contrast, hard-magnetic soft materials contain high-coercivity particles, typically neodymium-iron-boron (NdFeB), which retain a remanent magnetic state after pre-magnetisation. When the actuation field remains below the coercive field strength, this remanent state is approximately preserved, and deformation is driven mainly by the interaction between the programmed magnetic state and the externally applied field \cite{kim2018printing,zhao2019mechanics}. This distinction is central to constitutive modelling because the two material classes require different assumptions on the magnetic variables, coupling mechanisms and internal magnetic state.

\noindent The emergence of hard-magnetic soft materials has broadened the modelling landscape for MAPs. In soft-magnetic systems, a central challenge is to describe field-induced magnetisation, particle interactions and the dependence of the response on the evolving microstructure. In hard-magnetic systems, the remanent magnetic state is often prescribed, and the main task is to determine how this programmed state interacts with the applied magnetic field to generate bending, twisting, folding and other shape transformations. The ideal hard-magnetic soft-material model of Zhao et al. \cite{zhao2019mechanics} provides a particularly influential continuum description of this behaviour by introducing a residual magnetic flux density and an actuation-field regime below coercivity. This has provided a simple and effective basis for the model-guided design of shape-programmable magnetic soft structures.

\noindent The response of MAPs is also strongly influenced by particle distribution and curing history. If the composite is cured in the absence of a magnetic field, the particles are often approximately randomly distributed, and the material may be idealised as isotropic \cite{bell02,jolly1,varga1,bastola1}. If curing takes place under an applied magnetic field, the particles may form chain-like arrangements along preferred directions, giving rise to transversely isotropic or more general anisotropic behaviour \cite{bustamante1,danas1,hossain1,hossain2,saxena4}. This microstructural dependence directly affects the admissible energy functions, invariant sets, coupling terms and material parameters. It also motivates dispersed-chain and microstructurally informed models capable of representing imperfect particle alignment, chain distributions and particle-matrix interactions \cite{bustamante1,danas1,shariff5,saxena4}.

\noindent Constitutive models for MAPs must therefore account for several coupled effects, including field-induced stiffening, magnetostriction, magnetic saturation, anisotropic actuation, remanent magnetisation, hysteresis and rate-dependent polymer response \cite{ginder1,ginder4,bell02,danas1,saxena1,saxena3,haldar2021}. Since these materials often operate under large deformation, finite rotations and non-uniform magnetic fields, small-strain or purely phenomenological descriptions are insufficient for many practical applications. This has led to nonlinear finite-strain formulations based on continuum thermodynamics, invariant theory, variational principles, micromechanics and internal-variable approaches \cite{dorfmann3,dorfmann4,dorfmann5,kankanala1}.

\noindent The need for reliable finite-strain theories is especially clear in slender MAP structures such as membranes, beams, plates and shells, where large deformation, magnetic loading and geometric nonlinearity interact strongly. Recent work on magneto-mechanically coupled soft thin shells has shown that reduced-order theories must treat the magnetic field and the surrounding free space carefully, because Maxwell stresses and magnetic boundary conditions can influence the effective structural response \cite{ghosh2026shells}. Such developments complement constitutive modelling by showing how three-dimensional magnetoelastic laws enter lower-dimensional structural theories and computational frameworks.

\noindent The theoretical foundations of coupled magneto-mechanical modelling predate the modern literature on MAPs. Early works by Pao, Tiersten, Brown, and Eringen and Maugin established the continuum basis for electromagnetic interactions in deformable media \cite{pao1,tiersten1,brow66,eringen1,maugin1}. Building on this foundation, Dorfmann and Ogden developed a systematic finite-strain theory of magnetoelasticity based on a total energy function involving deformation and magnetic variables \cite{dorfmann3,dorfmann4,dorfmann5,dorfmann6}. In this framework, stresses and magnetic quantities are obtained directly from energy derivatives, which gives a compact and thermodynamically consistent basis for nonlinear boundary-value problems. Kankanala and Triantafyllidis further strengthened the variational basis of the subject by developing equivalent Eulerian and Lagrangian formulations for finitely strained magnetic particle-filled elastomers \cite{kankanala1}.

\noindent Since these foundational contributions, constitutive modelling of MAPs has developed in several directions. For isotropic elastic behaviour, the literature includes invariant-based finite-strain theories \cite{dorfmann3,dorfmann4,dorfmann5}, spectral formulations designed to improve physical interpretability \cite{bustamante4,shariff2,shariff3}, polyconvex approaches aimed at mathematical robustness \cite{silhavy2019,itskovKhiem2016}, and micromechanically motivated models that separate polymer-network and magnetic-particle contributions \cite{ethirajMiehe2016}. For anisotropic behaviour, key developments include transversely isotropic energy functions, experimentally calibrated finite-strain laws and dispersed-chain formulations that account for imperfect particle-chain alignment \cite{bustamante1,danas1,shariff5,saxena4}. Variational and mixed finite-element formulations have also enabled the treatment of non-homogeneous boundary-value problems and the inclusion of free-space magnetic effects in coupled computations \cite{bustamante5,vogel1,vogel4,pelteret1}.

\noindent Rate-dependence forms another important branch of constitutive modelling. Because the polymer matrix is intrinsically viscoelastic, purely elastic theories cannot fully describe transient, cyclic or relaxation phenomena. This has led to finite-strain magneto-viscoelastic models in which the mechanical and magnetic responses are decomposed into equilibrium and non-equilibrium parts \cite{saxena1,saxena3,nedjar2020}. More recent formulations have incorporated demagnetisation effects, field-induced Poynting-type responses, microstructure-sensitive dissipation mechanisms and experimental calibration under coupled loading \cite{haldar2021,garciaGonzalezHossain2021,wang2023mrevisco}. In parallel, homogenisation-based and multiscale approaches have become increasingly important for connecting macroscopic constitutive laws to particle morphology, chain formation and collective magnetic interactions \cite{galipeau2013,kalina2023review,nadzharyan2022review}.

\noindent Despite substantial progress, the literature remains distributed across several modelling traditions, including semi-empirical models, invariant-based continuum theories, spectral and anisotropic formulations, variational and polyconvex approaches, viscoelastic frameworks, microstructurally informed descriptions and hard-magnetic soft-material models. This diversity makes it difficult to compare assumptions, identify common constitutive structures and assess the range of validity of different formulations. Moreover, thermo-magneto-mechanical coupling remains comparatively less developed than purely magnetoelastic and magneto-viscoelastic theories, despite its importance for practical applications \cite{saber2026thermo}. A survey organised around constitutive structure is therefore useful for clarifying how the main modelling approaches are related and where further development is required.

\noindent The aim of this article is to provide a structured review of constitutive modelling for MAPs at finite strains. The emphasis is placed on constitutive variables, energy functions, invariant structures, thermodynamic restrictions, anisotropic extensions, hard-magnetic remanent-field formulations and dissipative mechanisms. The review covers the progression from early semi-empirical descriptions to nonlinear isotropic and anisotropic magnetoelastic models, spectral and variational formulations, polyconvex and micromechanically motivated frameworks, dispersed-chain models, hard-magnetic soft-material models and magneto-viscoelastic theories. Section~\ref{basiceqn} summarises the basic equations of nonlinear magneto-elasto-statics, including finite kinematics, balance laws, constitutive equations and representative magnetoelastic energy functions. Section~\ref{constvmodel} presents the main survey of constitutive models, organised into soft-magnetic and hard-magnetic soft-material frameworks, with further distinctions between isotropic, anisotropic, elastic, viscoelastic and remanent-field-based descriptions. Finally, Section~\ref{concl} summarises the main conclusions and outlines current challenges and future research directions.

\section{Nonlinear magneto-elasto-statics}\label{basiceqn}

\subsection{Kinematics}
As MAP solids generally experience large strains, it is necessary to distinguish between the reference configuration \(\mathcal{B}_0\) and the current configuration \(\mathcal{B}_t\). A material point \(\boldsymbol X \in \mathcal{B}_0\) is mapped to its spatial position \(\boldsymbol x \in \mathcal{B}_t\) through the deformation map
\[
\boldsymbol x=\boldsymbol\chi(\boldsymbol X).
\]
Throughout the manuscript, quantities defined in the reference configuration are denoted by uppercase letters, whereas quantities in the current configuration are represented by lowercase letters. The deformation gradient ($\boldsymbol{F}$) and its determinant ($J$) are defined as
\begin{equation}
\boldsymbol F := \text{Grad}\,\boldsymbol\chi,
\qquad
J := \det (\boldsymbol F) >0.
\end{equation}
The condition \(J>0\) guarantees local preservation of orientation and excludes physically inadmissible deformations.
The left and right Cauchy-Green deformation tensors, $\boldsymbol{b}$ and $\boldsymbol{C}$, respectively, are introduced as
\begin{equation}
\boldsymbol b := \boldsymbol F \boldsymbol F^{T},
\qquad
\boldsymbol C := \boldsymbol F^{T}\boldsymbol F.
\end{equation}

\subsection{Balance laws}

\subsubsection{Spatial configuration}
In the deformed configuration \(\mathcal{B}_t\), the magnetic induction \(\mathbbm b\) is related to the magnetic field \(\mathbbm h\) and the magnetization \(\mathbbm m\) through
\begin{equation}\label{defdspat}
\mathbbm b = \mu_0 \left(\mathbbm h + \mathbbm m\right)
\qquad \text{in } \mathcal{B}_t,
\end{equation}
where \(\mu_0\) denotes the permeability of vacuum. In free space, where \(\mathbbm m=\mathbf 0\), this reduces to \(\mathbbm b=\mu_0\mathbbm h\).

Under magnetostatic conditions, with vanishing free current density and time-independent electric displacement, Maxwell's equations reduce to
\begin{equation}\label{Maxwell_eq}
\text{curl}\,\mathbbm h = \mathbf 0,
\qquad
\text{div}\,\mathbbm b = 0,
\qquad \text{in } \mathcal{B}_t.
\end{equation}
Since \(\text{curl}\,\mathbbm h=\mathbf 0\), the magnetic field can be represented in terms of a scalar potential \(\varphi\) as
\begin{equation}\label{electric_field_def}
\mathbbm h = -\text{grad}\,\varphi .
\end{equation}

The interaction between the magnetic field and the material is described by the ponderomotive body force
\begin{equation}
\boldsymbol b_t^{\mathrm{pon}} := \mathbbm m \cdot \nabla \mathbbm b.
\end{equation}
This body force can be expressed as the divergence of a ponderomotive stress tensor,
\begin{equation}
\text{div}\,\boldsymbol\sigma^{\mathrm{pon}} = \boldsymbol b_t^{\mathrm{pon}},
\end{equation}
which may be decomposed into a magnetization part and a Maxwell part as
\begin{equation}\label{pondm}
\boldsymbol\sigma^{\mathrm{pon}}
=
\boldsymbol\sigma^{\mathrm{mag}}
+
\boldsymbol\sigma^{\mathrm{max}}.
\end{equation}
A convenient representation is
\begin{equation}\label{mageq}
\boldsymbol\sigma^{\mathrm{mag}}
=
(\mathbbm m\cdot \mathbbm b)\,\boldsymbol i
-
\mathbbm m\otimes \mathbbm b,
\qquad
\boldsymbol\sigma^{\mathrm{max}}
=
- M_t\,\boldsymbol i
+
\frac{1}{\mu_0}\,\mathbbm b\otimes \mathbbm b,
\end{equation}
where
\begin{equation}
M_t=\frac{1}{2\mu_0}\,\mathbbm b\cdot\mathbbm b,
\end{equation}
is the magnetic energy density per unit spatial volume and \(\boldsymbol i\) is the spatial identity tensor.

In the absence of matter, \(\mathbbm m=\mathbf 0\), and hence \(\boldsymbol\sigma^{\mathrm{mag}}=\mathbf 0\). In free space the Maxwell stress is divergence-free:
\begin{equation}\label{div_free_cond}
\text{div}\,\boldsymbol\sigma^{\mathrm{max}}=\mathbf 0.
\end{equation}

Including magnetic interactions, the balance of linear momentum takes the form
\begin{equation}\label{bolm_spatial}
\text{div}\,\boldsymbol\sigma
+
\boldsymbol b_t^{\mathrm{pon}}
+
\boldsymbol b_t
=
\text{div}\,\boldsymbol\sigma^{\mathrm{tot}}
+
\boldsymbol b_t
=
\mathbf 0
\qquad \text{in } \mathcal{B}_t,
\end{equation}
where \(\boldsymbol b_t\) denotes the mechanical body force and the total Cauchy stress is defined by
\begin{equation}
\boldsymbol\sigma^{\mathrm{tot}}
=
\boldsymbol\sigma
+
\boldsymbol\sigma^{\mathrm{pon}}.
\end{equation}

Let the boundary be decomposed as
\[
\partial\mathcal B_t=\partial\mathcal B_t^\chi\cup \partial\mathcal B_t^{t},
\qquad
\partial\mathcal B_t^\chi\cap \partial\mathcal B_t^{t}=\emptyset.
\]
On \(\partial\mathcal B_t^\chi\), the deformation is prescribed by
\begin{equation}
\boldsymbol\chi=\boldsymbol\chi^{\mathrm p}.
\end{equation}
On \(\partial\mathcal B_t^{t}\), prescribed tractions \(\boldsymbol t_t^{\mathrm p}\) lead to the boundary condition
\begin{equation}
[\![\boldsymbol\sigma^{\mathrm{tot}}]\!]\cdot \boldsymbol n
=
-\boldsymbol t_t^{\mathrm p},
\qquad \text{on } \partial\mathcal B_t^{t},
\end{equation}
where \(\boldsymbol n\) is the outward unit normal and the jump operator is defined as
\[
[\![\bullet]\!] = \{\bullet\}^{\mathrm{out}}-\{\bullet\}^{\mathrm{in}}.
\]

The jump conditions for the magnetic fields are
\begin{equation}
\boldsymbol n\cdot [\![\mathbbm b]\!] =0,
\qquad
\boldsymbol n\times [\![\mathbbm h]\!] = \hat{\mathbbm j}^{\,f},
\end{equation}
where \(\hat{\mathbbm j}^{\,f}\) is the free surface current density. If no free surface currents are present, then \(\hat{\mathbbm j}^{\,f}=\mathbf 0\), and from \eqref{electric_field_def} one obtains continuity of the magnetic scalar potential:
\begin{equation}
[\![\varphi]\!]=0.
\end{equation}

\subsubsection{Material configuration}
The magnetic quantities can be pulled back to the reference configuration \(\mathcal{B}_0\) according to
\begin{equation}
\mathbbm H = \mathbbm h\,\boldsymbol F,
\qquad
\mathbbm M = \mathbbm m\,\boldsymbol F,
\qquad
\mathbbm B = J\,\mathbbm b\,\boldsymbol F^{-T}.
\end{equation}
In material form, Maxwell's equations become
\begin{equation}\label{maxwell_mat}
\text{Curl}\,\mathbbm H = \mathbf 0,
\qquad
\text{Div}\,\mathbbm B = 0
\qquad \text{in } \mathcal{B}_0.
\end{equation}
Accordingly, the magnetic field may be expressed through the scalar potential in the reference configuration as
\begin{equation}
\mathbbm H = -\text{Grad}\,\varphi .
\end{equation}

The constitutive relation between \(\mathbbm B\), \(\mathbbm H\), and \(\mathbbm M\) in the material setting reads
\begin{equation}
\mathbbm B
=
J\mu_0\,\boldsymbol C^{-1}(\mathbbm H+\mathbbm M)
\qquad \text{in } \mathcal{B}_0.
\end{equation}
In free space this reduces to
\[
\mathbbm B = J\mu_0\,\boldsymbol C^{-1}\mathbbm H.
\]

The total Cauchy stress may be transformed to the first Piola-Kirchhoff and second Piola-Kirchhoff stresses as
\begin{equation}\label{Piola-Cauchy}
\boldsymbol P^{\mathrm{tot}}
=
J\,\boldsymbol\sigma^{\mathrm{tot}}\boldsymbol F^{-T},
\qquad
\boldsymbol S^{\mathrm{tot}}
=
J\,\boldsymbol F^{-1}\boldsymbol\sigma^{\mathrm{tot}}\boldsymbol F^{-T}.
\end{equation}
The total Piola stress is decomposed into mechanical and magnetic contributions:
\begin{equation}
\boldsymbol P^{\mathrm{tot}}
=
\boldsymbol P
+
\boldsymbol P^{\mathrm{pon}}
=
\boldsymbol P
+
\boldsymbol P^{\mathrm{mag}}
+
\boldsymbol P^{\mathrm{max}}.
\end{equation}
A corresponding material representation is
\begin{equation}\label{magmaxeq}
\boldsymbol P^{\mathrm{mag}}
=
(\mathbbm M\cdot\mathbbm B)\,\boldsymbol F^{-T}
-
\mathbbm m\otimes \mathbbm B,
\qquad
\boldsymbol P^{\mathrm{max}}
=
- M_0\,\boldsymbol F^{-T}
+
\frac{1}{\mu_0}\,\mathbbm b\otimes \mathbbm B,
\end{equation}
where
\begin{equation}
M_0=\frac{1}{2\mu_0}J^{-1}\,\mathbbm B\cdot(\boldsymbol C\,\mathbbm B)
\end{equation}
is the magnetic energy density per unit reference volume.

The balance of linear momentum in \(\mathcal{B}_0\) becomes
\begin{equation}
\text{Div}\,\boldsymbol P^{\mathrm{tot}}+\boldsymbol b_0=\mathbf 0
\qquad \text{in } \mathcal{B}_0,
\end{equation}
with traction boundary condition
\begin{equation}
[\![\boldsymbol P^{\mathrm{tot}}]\!]\cdot \boldsymbol N
=
-\boldsymbol t_0^{\mathrm p}
\qquad \text{on } \partial\mathcal B_0^{t},
\end{equation}
and, in free space,
\begin{equation}
\text{Div}\,\boldsymbol P^{\mathrm{max}}=\mathbf 0.
\end{equation}
The magnetic jump conditions in the reference configuration read
\begin{equation}
\boldsymbol N\cdot [\![\mathbbm B]\!]=0,
\qquad
\boldsymbol N\times [\![\mathbbm H]\!]=\hat{\mathbbm J}^{\,f},
\end{equation}
where \(\hat{\mathbbm J}^{\,f}\) denotes the free surface current density in the material description.

\section{Constitutive modelling of MAPs}\label{constvmodel}

\subsection{Strain energy functions}
\begin{figure}[H]
\centering
\footnotesize
\setlength{\tabcolsep}{2pt}
\resizebox{\textwidth}{!}{%
\begin{tikzpicture}[
    >=Stealth,
    line/.style={draw=black, line width=1.2pt, -{Stealth[length=4mm,width=2.5mm]}},
    titlebox/.style={
        rectangle, rounded corners, draw=black, thick,
        fill=gray!15, align=center, minimum width=7.4cm, minimum height=1.0cm
    },
    mainbox/.style={
        rectangle, rounded corners, draw=black, thick,
        fill=gray!10, align=center, text width=6.2cm, minimum height=1.0cm
    },
    catbox/.style={
        rectangle, rounded corners, draw=black, thick,
        fill=blue!10, align=center, text width=4.4cm, minimum height=0.9cm
    },
    subbox/.style={
        rectangle, rounded corners, draw=black, thick,
        fill=green!10, align=center, text width=4.2cm, minimum height=0.9cm
    },
    modelbox/.style={
        rectangle, rounded corners, draw=black, thick,
        fill=orange!10, align=left, text width=4.55cm, inner sep=6pt
    },
    hardbox/.style={
        rectangle, rounded corners, draw=black, thick,
        fill=orange!10, align=left, text width=7.4cm, inner sep=7pt
    }
]

\node[titlebox] (root) at (0,0) {Constitutive models for MAPs};

\node[mainbox] (soft) at (0,-1.7) {Models for soft-magnetic MAPs};
\node[mainbox] (hard) at (0,-11.0) {Models for hard-magnetic MAPs};

\draw[line] (root.south) -- (soft.north);

\draw[line] (root.east) -- ++(7.5,0) -- ++(0,-11.0) -- (hard.east);

\node[catbox] (iso) at (-5.3,-3.4) {Isotropic models};
\node[catbox] (aniso) at (5.3,-3.4) {Anisotropic models};

\draw[line] ([xshift=-1.8cm]soft.south) -- (iso.north);
\draw[line] ([xshift= 1.8cm]soft.south) -- (aniso.north);

\node[subbox] (isoe) at (-8.0,-5.2) {Elastic models};
\node[subbox] (isov) at (-2.6,-5.2) {Viscoelastic models};
\node[subbox] (anisoe) at (2.6,-5.2) {Elastic models};
\node[subbox] (anisov) at (8.0,-5.2) {Viscoelastic models};

\draw[line] ([xshift=-1.35cm]iso.south) -- (isoe.north);
\draw[line] ([xshift= 1.35cm]iso.south) -- (isov.north);
\draw[line] ([xshift=-1.35cm]aniso.south) -- (anisoe.north);
\draw[line] ([xshift= 1.35cm]aniso.south) -- (anisov.north);

\node[modelbox] (isoeM) at (-8.0,-8.25) {
$\bullet$ Jolly--Carlson--Mu\~noz semi-empirical model\\
$\bullet$ Dorfmann--Ogden model\\
$\bullet$ Steigmann membrane theory\\
$\bullet$ Kankanala--Triantafyllidis model\\
$\bullet$ Shariff spectral model\\
$\bullet$ Polyconvex/Itskov line\\
$\bullet$ Ethiraj--Miehe multiplicative model
};

\node[modelbox] (isovM) at (-2.6,-7.8) {
$\bullet$ Saxena--Hossain--Steinmann model\\
$\bullet$ Haldar model\\
$\bullet$ Nedjar model\\
$\bullet$ Garcia-Gonzalez--Hossain model
};

\node[modelbox] (anisoeM) at (2.6,-7.8) {
$\bullet$ Bustamante model\\
$\bullet$ Danas model\\
$\bullet$ Saxena--Pelteret--Steinmann dispersed-chain model\\
$\bullet$ Shariff spectral model
};

\node[modelbox] (anisovM) at (8.0,-7.8) {
$\bullet$ Saxena--Hossain--Steinmann transversely isotropic model
};

\draw[line] (isoe.south) -- (isoeM.north);
\draw[line] (isov.south) -- (isovM.north);
\draw[line] (anisoe.south) -- (anisoeM.north);
\draw[line] (anisov.south) -- (anisovM.north);

\node[hardbox] (hardM) at (0,-13.8) {
$\bullet$ Zhao--Kim--Chester--Sharma--Zhao model\\
$\bullet$ Mukherjee--Rambausek--Danas dissipative model\\
$\bullet$ Mukherjee--Danas unified dual formulation\\
$\bullet$ Danas--Reis stretch-independent magnetisation model\\
$\bullet$ Lin--Hooshmand-Ahoor--Bodelot--Danas foam model
};

\draw[line] (hard.south) -- (hardM.north);

\end{tikzpicture}%
}
\caption{Classification of constitutive models for MAPs discussed in Section~\ref{constvmodel}, with soft-magnetic soft materials separated into isotropic and anisotropic constitutive models, and hard-magnetic soft materials shown as a separate branch.}
\label{fig:constitutive_chart}
\end{figure}
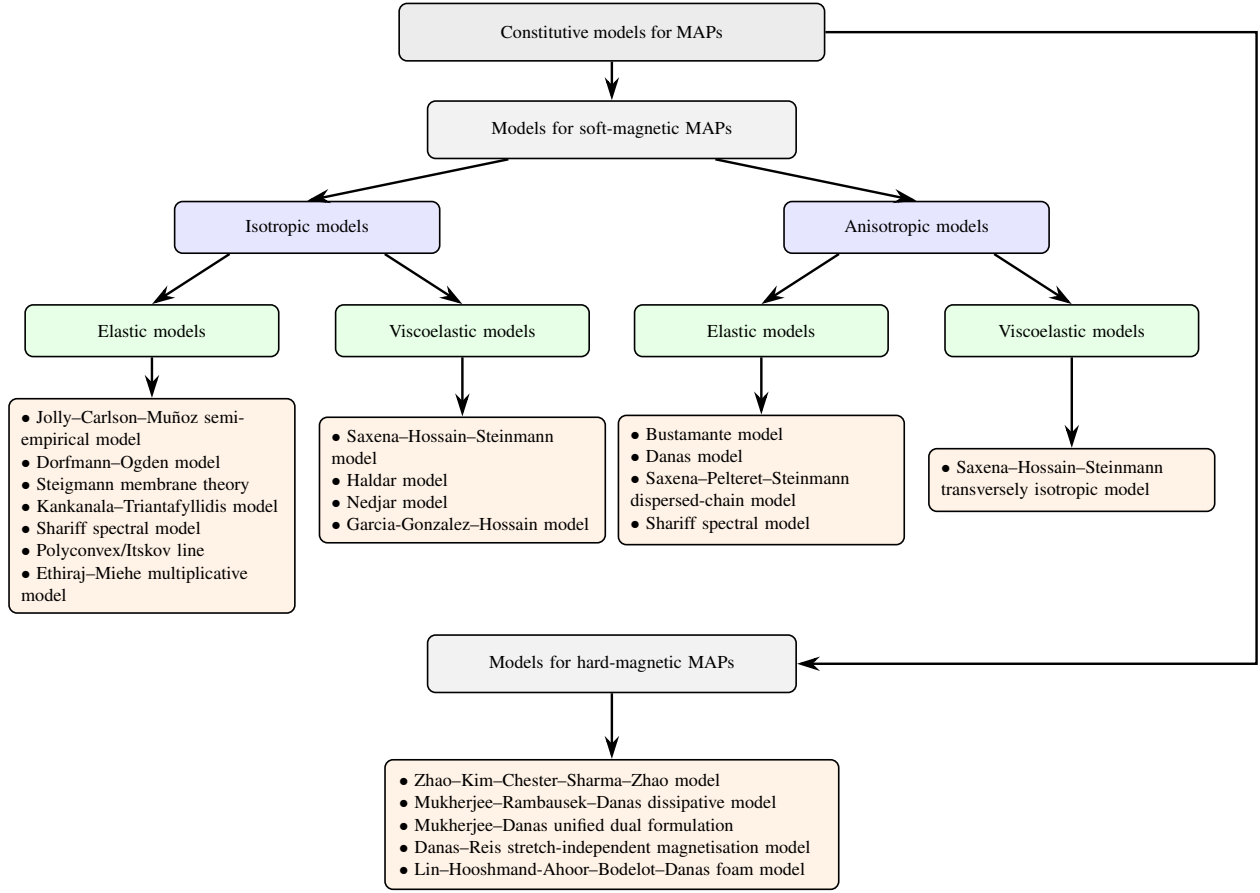

The constitutive models reviewed in this work are organised according to the classification shown in Figure~\ref{fig:constitutive_chart}. Before discussing each model in detail, the essential ingredients of such constitutive models are presented below.

Similar to strain energy functions for (passive) hyperelastic materials, the energy functions for MAPs are most conveniently expressed in terms of invariants formed from the right Cauchy-Green tensor \(\boldsymbol C\) and the material magnetic field \(\mathbbm H\). A standard set of invariants, for example,  is
\begin{equation}\label{definvar}
\begin{aligned}
I_1 &= \mathrm{tr}(\boldsymbol C), \qquad
&I_2 &= \tfrac12\Big[(\mathrm{tr}\,\boldsymbol C)^2-\mathrm{tr}(\boldsymbol C^2)\Big], \qquad
&I_3 &= \det(\boldsymbol C)=J^2, \\
I_4 &= \mathbbm H\cdot\mathbbm H, \qquad
&I_5 &= \mathbbm H\cdot(\boldsymbol C^{-1}\mathbbm H), \qquad
&I_6 &= \mathbbm H\cdot(\boldsymbol C^{-2}\mathbbm H).
\end{aligned}
\end{equation}
Hence, the typical strain energy function ($\Omega$) for MAPs may be written as a function of invariants as
\begin{equation}
\Omega=\Omega(I_1,I_2,I_3,I_4,I_5,I_6).
\end{equation}

If the magnetic contribution of the surrounding free space is neglected, then the total energy may be approximated directly by the material magnetoelastic energy, i.e.
\begin{equation}
\Omega(\boldsymbol F,\mathbbm H)\approx \Psi(\boldsymbol F,\mathbbm H).
\end{equation}

For a purely magnetoelastic material, the Helmholtz free energy per unit reference volume is taken as a function of the deformation gradient and the material magnetic field,
\begin{equation}\label{psi_mag}
\Psi=\Psi(\boldsymbol F,\mathbbm H).
\end{equation}
When the magnetic field is selected as the primary magnetic variable, it is convenient to introduce the complementary free-space magnetic energy
\begin{equation}\label{M0star}
M_0^{*}(\boldsymbol F,\mathbbm H)
:=
-\frac{1}{2}J\mu_0\,\mathbbm H\cdot\big[\boldsymbol C^{-1}\mathbbm H\big].
\end{equation}
The total magnetoelastic energy is then defined by
\begin{equation}\label{Omega_mag}
\Omega(\boldsymbol F,\mathbbm H)
=
\Psi(\boldsymbol F,\mathbbm H)
+
M_0^{*}(\boldsymbol F,\mathbbm H).
\end{equation}

In the absence of dissipation, the constitutive equations follow from the total energy as
\begin{equation}\label{const_rel_mag}
\boldsymbol P^{\mathrm{tot}}
=
\frac{\partial \Omega}{\partial \boldsymbol F},
\qquad
\mathbbm B
=
-
\frac{\partial \Omega}{\partial \mathbbm H}.
\end{equation}
If a formulation in terms of \(\boldsymbol C\) is preferred, the total second Piola-Kirchhoff stress is given by
\begin{equation}
\boldsymbol S^{\mathrm{tot}}
=
2\frac{\partial \Omega}{\partial \boldsymbol C}.
\end{equation}
The total Cauchy stress follows from the standard Piola transformation,
\begin{equation}
\boldsymbol\sigma^{\mathrm{tot}}
=
\frac{1}{J} \boldsymbol P^{\mathrm{tot}}\boldsymbol F^{T}.
\end{equation}

For incompressible magnetoelastic materials, the constraint \(J=1\) is enforced by a Lagrange multiplier \(p\). In that case,
\begin{equation}\label{incomp_const}
\boldsymbol P^{\mathrm{tot}}
=
\frac{\partial \Omega}{\partial \boldsymbol F}
-
p\,\boldsymbol F^{-T},
\qquad
\boldsymbol S^{\mathrm{tot}}
=
2\frac{\partial \Omega}{\partial \boldsymbol C}
-
p\,\boldsymbol C^{-1}.
\end{equation}

\subsection{Constitutive models for soft-magnetic MAPs}

For soft-magnetic MAPs, the development proceeds from early semi-empirical descriptions of field-induced stiffening to nonlinear isotropic finite-strain theories, and then to anisotropic, dispersed-chain, microstructural and rate-dependent formulations \cite{jolly1,dorfmann3,dorfmann4,dorfmann5,kankanala1,bustamante1,danas1,saxena4}. Within this classification, the model of Jolly, Carlson and Mu\~noz provides an early mechanistic description, while the Dorfmann--Ogden and Kankanala--Triantafyllidis frameworks provide the main continuum foundations for isotropic finite-strain magnetoelasticity \cite{jolly1,dorfmann3,dorfmann4,dorfmann5,dorfmann6,kankanala1}. Later contributions extend these ideas to anisotropy, spectral representations, dispersed chain-like microstructures, micromechanically motivated energy functions and dissipative response \cite{bustamante1,danas1,saxena1,saxena3,saxena4,shariff5,ethirajMiehe2016,nedjar2020,haldar2021,garciaGonzalezHossain2021}.

\subsubsection{Isotropic elastic constitutive models}\label{constvmodeliso}

\paragraph{\textbf{Jolly--Carlson--Munoz semi-empirical model}}
A historically important precursor to nonlinear finite-strain continuum theories is the semi-empirical model proposed by Jolly, Carlson and Munoz \cite{jolly1}. The model is quasi-static and one-dimensional, and is therefore not a general finite-strain tensorial constitutive law. It considers chains of magnetically permeable spherical particles embedded in a non-magnetic medium and subjected to a magnetic field approximately parallel to the particle chains. The field-induced mechanical response is attributed to magnetic dipole interactions between neighbouring particles in a chain, while the zero-field viscoelastic response of the composite is assumed to be superposed separately and is not explicitly modelled.

For two neighbouring particles idealised as identical magnetic dipoles of magnitude $m_{\mathrm d}$, separated by a distance $r_0$ in the undeformed chain direction and sheared by a transverse displacement $x$, the scalar shear strain is defined as
\begin{equation}
\gamma=\frac{x}{r_0}.
\end{equation}
The magnetic interaction energy between the two dipoles is written as
\begin{equation}
E_{12}
=
\frac{(\gamma^2-2)m_{\mathrm d}^2}
{4\pi\mu_1\mu_0 r_0^3(1+\gamma^2)^{5/2}},
\end{equation}
where $\mu_1$ is the relative permeability of the surrounding medium. By averaging the pair interaction energy over aligned particle chains, the corresponding magnetic energy density becomes
\begin{equation}
W
=
\frac{3\phi(\gamma^2-2)m_{\mathrm d}^2}
{2\pi^2\mu_1\mu_0 d^3 r_0^3(1+\gamma^2)^{5/2}},
\end{equation}
where $\phi$ is the particle volume fraction and $d$ is the particle diameter.

The magnetic-field-induced shear stress is obtained by differentiating the interaction energy density with respect to the scalar shear strain:
\begin{equation}
\tau
=\frac{\partial W}{\partial \gamma}=
\frac{9\phi\gamma(4-\gamma^2)m_{\mathrm d}^2}
{2\pi^2\mu_1\mu_0 d^3 r_0^3(1+\gamma^2)^{7/2}}.
\end{equation}
Introducing the average particle polarization $J_p$, 
and setting $h=\dfrac{r_0}{d}$.
thus, the field-induced stress is proportional to the square of the particle polarization and inversely proportional to the cube of the interparticle spacing parameter $h$.
The yield stress of the idealized particle chain is obtained by maximizing the above stress with respect to $\gamma$, giving the following:
\begin{equation}
\gamma_y
=
\frac{\sqrt{27-\sqrt{665}}}{2^{3/2}}
\simeq 0.389,
\end{equation}
and
\begin{equation}
\tau_y
=
\frac{0.1143,\phi J_p^2}
{\mu_1\mu_0 h^3}.
\end{equation}
For small shear strains, ($\gamma<0.1$), the field-induced stress reduces to
\begin{equation}
\tau
\simeq
\frac{\phi\gamma J_p^2}
{2\mu_1\mu_0 h^3},
\end{equation}
so that the pre-yield field-induced shear modulus of the particle network is
\begin{equation}
G_{\mathrm{mag}}
\simeq
\frac{\phi J_p^2}
{2\mu_1\mu_0 h^3}.
\end{equation}
If the zero-field shear modulus of the elastomeric composite is denoted by $G_0$, then the small-strain effective shear modulus may be written in additive form.
\begin{equation}
G_{\mathrm{eff}}
\simeq
G_0+G_{\mathrm{mag}},
\end{equation}
with the understanding that $G_0$ represents the non-magnetic mechanical response, whereas $G_{\mathrm{mag}}$ is the field-induced contribution predicted by the particle-chain model.

The model also relates the particle polarization $J_p$ to the average composite magnetic flux density by introducing an approximate magnetic saturation mechanism. In this part of the theory, saturation is assumed to begin in the polar regions of each particle and to progress with increasing applied field. The resulting model is semi-empirical because a fitting parameter is required to account for magnetic interactions not captured by the idealised one-dimensional chain description. Therefore, the Jolly--Carlson--Munoz model should be understood as a mechanistically motivated, small-strain, particle-chain model for field-induced stiffening, rather than as a general thermodynamically consistent finite-strain magnetoelastic energy function.

\paragraph{\textbf{Dorfmann--Ogden model}}
The finite-strain constitutive treatment of isotropic magnetoelastic elastomers is strongly rooted in the work of Dorfmann and Ogden \cite{dorfmann3,dorfmann4,dorfmann5,dorfmann6}. Their formulation describes the coupled magneto-mechanical response through a total energy density depending on the deformation and a magnetic variable. Objectivity and isotropy reduce this dependence to a scalar function of a finite set of invariants. Using the material magnetic field \(\mathbbm H\) as the independent magnetic variable, the energy may be written as
\begin{equation*}
\Omega
=
\Omega(\boldsymbol F,\mathbbm H)
=
\Omega(I_1,I_2,I_3,I_4,I_5,I_6),
\end{equation*}
where, $I_4$, $I_5$, and $I_6$ are given from the equation set \eqref{definvar}.
 In the incompressible case, \(J=1\), so that \(I_3=1\), and the constitutive dependence reduces to
\begin{equation}
\Omega=\Omega(I_1,I_2,I_4,I_5,I_6).
\end{equation}

A simple representative isotropic magnetoelastic energy within this class is
\begin{equation}
\Omega
=
\frac{\mu}{2}(I_1-3)
+
\frac{c_4}{2}I_4
+
\frac{c_5}{2}I_5
+
\frac{c_6}{2}I_6,
\end{equation}
where \(\mu\) is the shear modulus, \(c_4\) is a magnetic material parameter, and \(c_5\) and \(c_6\) are magneto-mechanical coupling parameters. The first term represents the neo-Hookean elastic response, the second term represents magnetic storage, and the last two terms introduce deformation-dependent magnetic coupling. More elaborate versions may replace the neo-Hookean elastic term by Mooney--Rivlin, Ogden, Gent, or Arruda--Boyce energies, and may include nonlinear magnetic terms to account for magnetic saturation.

The constitutive equations are obtained directly from the total energy. For an incompressible material with \(\mathbbm H\) chosen as the independent magnetic variable, they read from equations \eqref{const_rel_mag} and \eqref{incomp_const}, respectively.
%
The strength of the Dorfmann--Ogden model lies in its compact energetic structure and thermodynamic consistency. Since stresses and magnetic quantities are obtained by differentiating a single total energy function, the framework is well suited for constitutive specialization and for the analysis of nonlinear boundary-value problems \cite{dorfmann3,dorfmann4,dorfmann5,dorfmann6}. It should be noted, however, that the Dorfmann--Ogden formulation is mainly a general theoretical framework, not a single experimentally calibrated material model. Its invariant structure provides the basis for constructing admissible energy functions, but any particular choice of \(\Omega(I_1,I_2,I_4,I_5,I_6)\) must be calibrated separately for the material under consideration. Thus, its validation is largely indirect, through its use in analytical and numerical boundary-value studies \cite{dorfmann6,bustamante5,bustamante7,pelteret1} and through later constitutive models that were calibrated against or compared with experimental magneto-mechanical data \cite{kankanala1,danas1,saxena4,salasBustamante2015}. In this sense, it has become a canonical reference formulation for subsequent isotropic, anisotropic, spectral, variational, and microstructurally enriched theories of finite magnetoelasticity \cite{kankanala1,bustamante4,shariff5,danas1,saxena4}.

\paragraph{\textbf{Steigmann membrane theory}}
Steigmann's magnetoelastic membrane theory provides a reduced-dimensional
framework for thin magnetoelastic bodies \cite{steigmann1}. It is derived from
a variational three-dimensional magnetoelastic formulation and is then used to
motivate a direct membrane theory. Thus, its main contribution is not a new
bulk constitutive law, but the identification of an admissible surface-energy
structure for magnetoelastic membranes.

In the direct membrane setting, the surface energy may be written as
\begin{equation}
\mathcal W
=
\mathcal W(\boldsymbol C_s,\mathbbm H_s,\boldsymbol A_s,
J_s\,\boldsymbol n\cdot\boldsymbol a),
\qquad
\boldsymbol A_s=\boldsymbol F_s^{T}\boldsymbol a ,
\end{equation}
where \(\boldsymbol C_s=\boldsymbol F_s^{T}\boldsymbol F_s\) is the surface
right Cauchy--Green tensor, \(\mathbbm H_s\) is the material magnetic field on
the membrane surface, \(\boldsymbol a\) is associated with the applied magnetic
loading, and \(\boldsymbol A_s\) is its pull-back to the reference membrane
surface. Here, \(J_s\) denotes the surface Jacobian, or local area stretch, so
that
\begin{equation}
da = J_s\,dA .
\end{equation}
The scalar \(J_s\,\boldsymbol n\cdot\boldsymbol a\) therefore represents the
area-scaled normal contribution of the applied magnetic loading.

For an isotropic membrane, the energy can be expressed as
\begin{equation}
\mathcal W
=
\widehat{\mathcal W}(i_1,\ldots,i_9),
\end{equation}
with
\begin{equation}
\begin{aligned}
&i_1=\boldsymbol I_s:\boldsymbol C_s,
\qquad
i_2=\det\boldsymbol C_s,
\qquad
i_3=\boldsymbol C_s:(\mathbbm H_s\otimes\mathbbm H_s),\\
&i_4=\boldsymbol C_s:(\boldsymbol A_s\otimes\boldsymbol A_s),
\qquad
i_5=\boldsymbol C_s:(\boldsymbol A_s\otimes\mathbbm H_s),
\qquad
i_6=\mathbbm H_s\cdot\mathbbm H_s,\\
&i_7=\boldsymbol A_s\cdot\boldsymbol A_s,
\qquad
i_8=\boldsymbol A_s\cdot\mathbbm H_s,
\qquad
i_9=(J_s\,\boldsymbol n\cdot\boldsymbol a)^2 .
\end{aligned}
\end{equation}

\paragraph{\textbf{Kankanala--Triantafyllidis variational model}}
A distinct finite-strain framework for bulk isotropic magnetoelasticity was developed by Kankanala and Triantafyllidis \cite{kankanala1}. Their formulation is important because it establishes the equivalence between two complementary descriptions: an Eulerian approach based on balance laws and thermodynamics, and a Lagrangian approach based on minimization of a total potential-energy functional. In this model, the specific Helmholtz free energy is taken to depend on the deformation and magnetization. In a Lagrangian representation, it may be written as
\begin{equation}
\Omega=\psi(\boldsymbol C,\mathbbm M),
\end{equation}
where \(\mathbbm M\) denotes the magnetization variable. For an isotropic material, objectivity reduces the dependence of the energy to a scalar function of mechanical, magnetic, and magneto-mechanical invariants. In the incompressible isotropic case, this may be written as
\begin{equation}
\psi=\hat{\psi}(I_1,I_2,J_1,J_2,J_3),
\end{equation}
where
\begin{equation}
\begin{aligned}
&J_1=\mathbbm M\cdot\mathbbm M,
\qquad
J_2=\mathbbm M\cdot\boldsymbol C\mathbbm M,
\qquad
J_3=\mathbbm M\cdot\boldsymbol C^2\mathbbm M .
\end{aligned}
\end{equation}

The magnetic constitutive relation is obtained from
\begin{equation}
\mu_0\mathbbm h
=
\frac{\partial \psi}{\partial \mathbbm M},
\end{equation}
where \(\mathbbm h\) is the magnetic field and \(\mu_0\) is the permeability of free space. The total Cauchy stress is expressed as
\begin{equation}
\boldsymbol\sigma^{\mathrm{tot}}
=
\rho
\left[
2\boldsymbol F
\frac{\partial \psi}{\partial \boldsymbol C}
\boldsymbol F^T
+
\mu_0
\left(
\mathbbm M\otimes\mathbbm h
+
\mathbbm h\otimes\mathbbm M
\right)
\right]
+
\mu_0
\left[
\mathbbm h\otimes\mathbbm h
-
\frac{1}{2}
(\mathbbm h\cdot\mathbbm h)\boldsymbol i
\right],
\end{equation}
where \(\rho\) is the current mass density. This stress contains the elastic, magnetization, and Maxwell-type magnetic contributions in a single total stress tensor. A representative fitted energy proposed for incompressible magneto-active elastomers is
\begin{equation}
\begin{aligned}
\rho_0\psi
&=
\frac{\mu}{2}
\left[
\left(C_{10}+C_{11}\frac{J_1}{M_s^2}\right)(I_1-3)
+
\left(C_{20}+C_{21}\frac{J_1}{M_s^2}\right)(I_2-3)
\right.\\
&\hspace{2.2cm}\left.
+
C_{01}\frac{J_1}{M_s^2}
+
C_{02}\frac{J_2}{M_s^2}
+
C_{01}^{*}
\left\{
\cosh\left(\frac{J_1}{M_s^2}\right)-1
\right\}
\right].
\end{aligned}
\end{equation}
Here, \(\rho_0\) is the reference mass density, \(M_s\) is the saturation magnetization, and \(C_{10}\), \(C_{11}\), \(C_{20}\), \(C_{21}\), \(C_{01}\), \(C_{02}\), and \(C_{01}^{*}\) are material parameters. The first two terms describe the elastic response with magnetization-dependent stiffness, the \(J_1\)-dependent terms represent magnetic storage and saturation-type effects, and the \(J_2\) term introduces deformation-dependent magneto-mechanical coupling.

The main contribution of the Kankanala--Triantafyllidis model is that it places finite-strain magneto-active elastomer modelling on a clear variational and thermodynamic basis. By combining physically meaningful magnetic variables with a minimization framework, the model provides both a practical constitutive law for magneto-active elastomers and a mathematically robust foundation for nonlinear analysis and numerical implementation.

\paragraph{\textbf{Shariff spectral model}}
Bustamante and Shariff proposed a spectral reformulation of nonlinear isotropic magnetoelasticity in which the constitutive variables are expressed in terms of principal stretches and magnetic-field projections rather than only through the classical invariant basis \cite{bustamante4}. The aim of this approach is to improve the physical interpretability of the constitutive law while retaining the finite-strain structure of the general isotropic theory. Let the spectral decomposition of the right Cauchy--Green tensor be
\begin{equation}
\boldsymbol C
=
\sum_{a=1}^{3}
\lambda_a^2\boldsymbol N_a\otimes\boldsymbol N_a,
\end{equation}
where \(\lambda_a\) are the principal stretches and \(\boldsymbol N_a\) are the corresponding material principal directions. The material magnetic field is projected onto the same principal basis as
\begin{equation}
H_a=\mathbbm H\cdot\boldsymbol N_a,
\qquad a=1,2,3.
\end{equation}
The magnetoelastic energy may then be written in the spectral form
\begin{equation}
\Omega
=
\widehat{\Omega}_{\mathrm{Sh}}
(\lambda_1,\lambda_2,\lambda_3,H_1,H_2,H_3),
\end{equation}
where \(\widehat{\Omega}_{\mathrm{Sh}}\) denotes the spectral representation of the magnetoelastic energy in the Shariff formulation. A simple representative example is
\begin{equation}
\widehat{\Omega}_{\mathrm{Sh}}
=
\sum_{a=1}^{3}
\frac{\mu_a}{2}(\lambda_a^2-1-\ln\lambda_a^2)
+
\frac{1}{2}\sum_{a=1}^{3}\alpha_a H_a^2
+
\frac{1}{2}\sum_{a=1}^{3}\beta_a\lambda_a^2H_a^2,
\end{equation}
where \(\mu_a\), \(\alpha_a\), and \(\beta_a\) are material parameters. The first term represents the elastic contribution written in terms of principal stretches, the second term represents the purely magnetic contribution, and the third term introduces deformation-dependent magneto-mechanical coupling. For incompressible materials, the stretches satisfy
\begin{equation}
\lambda_1\lambda_2\lambda_3=1.
\end{equation}

The principal constitutive equations follow by differentiation of the spectral energy:
\begin{equation}
P_a
=
\frac{\partial\widehat{\Omega}_{\mathrm{Sh}}}{\partial\lambda_a}
-
p\lambda_a^{-1},
\qquad
B_a
=
-
\frac{\partial\widehat{\Omega}_{\mathrm{Sh}}}{\partial H_a},
\qquad a=1,2,3,
\end{equation}
where \(P_a\) are the principal nominal stress components, \(B_a\) are the components of the material magnetic induction in the principal basis. This spectral representation gives the model a clearer geometric interpretation than the conventional invariant form involving \(I_4\), \(I_5\), and \(I_6\). It is therefore useful for parameter identification and for interpreting how individual deformation modes interact with the magnetic field.

\paragraph{\textbf{Polyconvex/Itskov model}}
The polyconvex approach to constitutive modelling is motivated primarily by mathematical robustness. Its purpose is to construct free-energy functions for coupled electro-magneto-elastic materials with convexity properties that are useful for existence theory, stability analysis, and reliable numerical implementation. In this line, Itskov and co-workers proposed polyconvex formulations for electro- and magneto-active elastomers, while related mathematical developments provide systematic restrictions on admissible coupled-field energy functions \cite{itskovKhiem2016,silhavy2019}.

The polyconvex stored-energy function is denoted here by \(\Omega_{\mathrm{pc}}\). A representative polyconvex magnetoelastic energy can be written as a convex function of an enlarged set of arguments,
\begin{equation}
\Omega(\boldsymbol F,\mathbbm H)
=
\Omega_{\mathrm{pc}}
\left(
\boldsymbol F,
\operatorname{cof}\boldsymbol F,
J,
\mathbbm H,
\boldsymbol F^{-T}\mathbbm H
\right).
\end{equation}
Here, \(\operatorname{cof}\boldsymbol F\) denotes the cofactor of the deformation gradient, defined by
\begin{equation}
\operatorname{cof}\boldsymbol F
=
J\boldsymbol F^{-T}.
\end{equation}
Thus, the energy is not expressed only through \(\boldsymbol F\), but also through its cofactor and determinant, which are the standard arguments used in polyconvex finite elasticity. The additional argument \(\boldsymbol F^{-T}\mathbbm H\) introduces deformation-dependent coupling with the material magnetic field.

A simple admissible example can be written as
\begin{equation}
\Omega
=
\frac{\mu}{2}\left(I_1-3-2\ln J\right)
+
\frac{\kappa}{2}(J-1)^2
+
\frac{c_4}{2}\mathbbm H\cdot\mathbbm H
+
\frac{c_5}{2}(\boldsymbol F^{-T}\mathbbm H)\cdot(\boldsymbol F^{-T}\mathbbm H).
\end{equation}
 Equivalently, the coupling term may be written as
\begin{equation*}
(\boldsymbol F^{-T}\mathbbm H)\cdot(\boldsymbol F^{-T}\mathbbm H)
=
\mathbbm H\cdot\boldsymbol C^{-1}\mathbbm H,
\end{equation*}
which makes its connection with the usual invariant-based magnetoelastic coupling clear.

The constitutive equations retain the usual energetic form as given by equations \eqref{const_rel_mag} and \eqref{incomp_const}.
The polyconvex line of modelling therefore complements classical invariant-based theories by providing a mathematically disciplined framework for finite-strain magnetoelastic constitutive design.

\paragraph{\textbf{Ethiraj--Miehe multiplicative model}}
Ethiraj and Miehe proposed a micromechanically motivated finite-strain model for isotropic magnetoactive elastomers \cite{ethirajMiehe2016}. In contrast to invariant-based magnetoelastic formulations, the magneto-mechanical coupling is introduced through a multiplicative kinematic split of the deformation gradient. The formulation uses the spatial magnetization as the primary magnetic variable and allows standard polymer-network models to be incorporated in a modular way.
 The material energy density is written here as
\begin{equation}
\Omega
=
\Omega_{\mathrm{mag}}(\mathbbm m)
+
\Omega_{\mathrm{mec}}(\boldsymbol F,\mathbbm m),
\end{equation}
where \(\Omega_{\mathrm{mag}}\) represents the magnetic energy of the magnetized rigid particles and \(\Omega_{\mathrm{mec}}\) represents the elastic energy of the deformable polymer network.

For nearly incompressible response, the deformation gradient is decomposed into volumetric and isochoric parts,
\begin{equation}
\bar{\boldsymbol F}=J^{-1/3}\boldsymbol F .
\end{equation}
The energy is then expressed as
\begin{equation}
\Omega(\boldsymbol F,\mathbbm m)
=
\Omega_{\mathrm{vol}}(J)
+
\Omega_{\mathrm{mag}}(\mathbbm m)
+
\Omega_{\mathrm{net}}(\boldsymbol F_{\mathrm{net}}),
\end{equation}
where \(\Omega_{\mathrm{vol}}(J)\) is the volumetric contribution and \(\Omega_{\mathrm{net}}\) is the polymer-network energy. A standard volumetric penalty may be taken as
\begin{equation}
\Omega_{\mathrm{vol}}(J)
=
\frac{\kappa}{2}(J-1)^2,
\end{equation}
with \(\kappa\) denoting the bulk modulus.

The central assumption of the model is a left multiplicative split of the isochoric deformation,
\begin{equation}
\bar{\boldsymbol F}
=
\boldsymbol E^{-1}(\mathbbm m)\boldsymbol F_{\mathrm{net}},
\qquad
\boldsymbol F_{\mathrm{net}}
=
\boldsymbol E(\mathbbm m)\bar{\boldsymbol F}.
\end{equation}
Here, \(\boldsymbol E^{-1}(\mathbbm m)\) represents the local magnetostrictive stretch and \(\boldsymbol F_{\mathrm{net}}\) is the stress-producing deformation of the polymer network. Thus,
\begin{equation}
\Omega_{\mathrm{mec}}(\boldsymbol F,\mathbbm m)
=
\Omega_{\mathrm{net}}(\boldsymbol F_{\mathrm{net}}).
\end{equation}

The magnetostrictive stretch is assumed to be aligned with the spatial magnetization:
\begin{equation}
\boldsymbol E^{-1}(\mathbbm m)
=
\boldsymbol i
+
c\,\mathbbm m\otimes\mathbbm m,
\end{equation}
where \(c\geq 0\) is a magnetostrictive material parameter and \(\boldsymbol i\) is the spatial identity tensor. Since this stretch is not volume preserving, its isochoric form is used in the network deformation. Defining
\begin{equation}
D
=
\det[\boldsymbol E^{-1}(\mathbbm m)]
=
1+c\,\mathbbm m\cdot\mathbbm m,
\end{equation}
one obtains
\begin{equation}
\boldsymbol E(\mathbbm m)
=
D^{1/3}
\left[
\boldsymbol i
-
\frac{c}{D}\mathbbm m\otimes\mathbbm m
\right].
\end{equation}
This construction induces a magnetization-dependent spatial metric,
\begin{equation}
\boldsymbol g(\mathbbm m)
=
\boldsymbol E^{T}(\mathbbm m)\boldsymbol i\,\boldsymbol E(\mathbbm m),
\end{equation}
which maps an initially isotropic polymer-chain network into a magnetization-dependent anisotropic network response.

The network right Cauchy--Green tensor and its first invariant are
\begin{equation}
\boldsymbol C_{\mathrm{net}}
=
\boldsymbol F_{\mathrm{net}}^{T}\boldsymbol F_{\mathrm{net}},
\qquad
I_1^{\mathrm{net}}
=
\mathrm{tr}(\boldsymbol C_{\mathrm{net}}).
\end{equation}
For a Gaussian-chain network, one may write
\begin{equation}
\Omega_{\mathrm{net}}
=
\frac{\mu}{2}
\left(
I_1^{\mathrm{net}}-3
\right),
\end{equation}
where \(\mu\) is the shear modulus of the polymer network. A finite-chain extensibility version can be introduced through an Arruda--Boyce-type expansion,
\begin{equation}
\Omega_{\mathrm{net}}
=
\mu
\sum_{k=1}^{K}
\frac{C_k}{N^{k-1}}
\left[
\left(I_1^{\mathrm{net}}\right)^k-3^k
\right],
\end{equation}
where \(N\) is the number of chain segments and \(C_k\) are the Arruda--Boyce coefficients.

The formulation also admits a microsphere representation:
\begin{equation}
\Omega_{\mathrm{net}}
=
\frac{1}{|\mathcal S|}
\int_{\mathcal S}
\psi_{\mathrm{chain}}
\left(
\lambda_{\mathrm{net}}(\boldsymbol r)
\right)
\,\mathrm dA,
\end{equation}
where \(\mathcal S\) is the unit sphere, \(\boldsymbol r\) is a unit chain direction, and
\begin{equation}
\lambda_{\mathrm{net}}(\boldsymbol r)
=
\sqrt{
\boldsymbol r\cdot
\boldsymbol C_{\mathrm{net}}\boldsymbol r
}.
\end{equation}
The single-chain energy \(\psi_{\mathrm{chain}}\) may be chosen from Gaussian, Langevin, or finite-extensibility-based chain statistics.

The magnetic particle contribution may be represented by a quadratic law,
\begin{equation}
\Omega_{\mathrm{mag}}(\mathbbm m)
=
\frac{C_{\mathrm m}}{2}
\mathbbm m\cdot\mathbbm m,
\end{equation}
where \(C_{\mathrm m}\) is a magnetic material parameter. To account for magnetic saturation, Ethiraj and Miehe also considered a statistical magnetic-particle kernel. In scalar form, with \(m=\|\mathbbm m\|\), this may be written as
\begin{equation}
\psi_{\mathrm{mag}}(m)
=
\frac{n k_{\mathrm B}\theta}{2}
\left[
\frac{m}{m_s}
\ln\left(
\frac{1+m/m_s}{1-m/m_s}
\right)
+
\ln\left(
1-\frac{m^2}{m_s^2}
\right)
\right],
\end{equation}
where \(m_s\) is the saturation magnetization and \(n\) is the number of magnetic dipoles per unit mass. For moderate magnetization, a truncated approximation is
\begin{equation}
\psi_{\mathrm{mag}}(m)
\approx
n k_{\mathrm B}\theta
\left[
\left(\frac{m}{m_s}\right)^2
+
\frac{5}{12}
\left(\frac{m}{m_s}\right)^4
\right].
\end{equation}

The constitutive quantities follow by differentiating the energy. The material nominal stress may be written from equation \eqref{const_rel_mag}
with the derivative evaluated through the chain,
\[
\boldsymbol F
\mapsto
\bar{\boldsymbol F}
\mapsto
\boldsymbol F_{\mathrm{net}}.
\]
The magnetic constitutive relation contains the direct magnetic-particle contribution and the additional contribution induced by the dependence of \(\boldsymbol F_{\mathrm{net}}\) on \(\mathbbm m\). For the material:
\begin{equation}
\mu_0\mathbbm h
=
\frac{\partial\Omega_{\mathrm{mag}}}{\partial\mathbbm m}
+
\frac{\partial\Omega_{\mathrm{net}}}{\partial\boldsymbol F_{\mathrm{net}}}
:
\frac{\partial\boldsymbol F_{\mathrm{net}}}{\partial\mathbbm m}.
\end{equation}

The main significance of the Ethiraj--Miehe model is its modular microstructural structure. The magnetic-particle energy and the polymer-network energy can be selected independently, while the multiplicative split transfers the effect of spatial magnetization directly into the network kinematics. The model is therefore best understood as an isotropic micromechanical magnetoelastic framework for magneto-active elastomers, rather than as a conventional phenomenological invariant-based energy.

\subsubsection{Isotropic viscoelastic constitutive models}\label{constvmodelisoviscoelas}

The preceding elastic formulations provide a useful basis for describing field-dependent deformation and stiffness changes in MAPs. Nevertheless, many MAPs also exhibit pronounced rate-dependent effects because the elastomeric matrix is intrinsically viscoelastic, while magnetic loading can introduce additional dissipation through particle interactions, field-dependent microstructural rearrangement and, in soft-magnetic systems, the evolution of the magnetic state \cite{bose2,saxena1}. Early experimental and semi-empirical studies reported dynamic modulus enhancement, relaxation, hysteresis and magnetostrictive deformation, showing that purely elastic constitutive models are generally insufficient for transient or cyclic loading conditions \cite{jolly1,ginder1,ginder4,bell02}.

The semi-empirical model of Jolly {et al.} \cite{jolly1} was historically important because it recognised the rate-dependent character of magnetic particle-filled elastomers and interpreted the magnetic contribution through idealised particle-chain interactions. However, that description remained quasi-static, essentially one-dimensional and outside a general finite-strain thermodynamic framework. The high-frequency measurements of Ginder {et al.} \cite{ginder4} further showed substantial field-induced changes in dynamic shear modulus, providing additional motivation for thermodynamically consistent internal-variable formulations.

In the models reviewed below, the energy is generally decomposed into an equilibrium contribution and one or more non-equilibrium contributions. When a generalized Maxwell-type representation is used, \(\alpha=1,\ldots,N\) denotes the dissipative-branch index, and \(N\) is the total number of branches. The equilibrium shear modulus is denoted by \(\mu_{\infty}\), while \(\mu_{\alpha}\) denotes the shear modulus associated with the \(\alpha\)-th non-equilibrium branch. The parameters \(a_{\infty}\), \(a_{\alpha}\), \(b_{\infty}\) and \(b_{\alpha}\) are used as representative equilibrium and branch-level magnetic or magneto-mechanical coupling constants when a unified review notation is helpful. The original papers may use different symbols for analogous constants.

\paragraph{\textbf{Saxena--Hossain--Steinmann isotropic magneto-viscoelastic model}}
Saxena, Hossain and Steinmann proposed one of the foundational finite-strain constitutive theories for isotropic magneto-viscoelasticity \cite{saxena1}. The model accounts for two dissipative mechanisms: mechanical viscoelasticity of the polymeric matrix and magnetic relaxation associated with the evolution of the magnetic induction. For each non-equilibrium branch, the deformation gradient is multiplicatively decomposed, while the magnetic induction is additively decomposed as
\begin{equation}\label{def_grad_vis_decomp}
\boldsymbol F
=
\boldsymbol F_{\mathrm e\alpha}
\boldsymbol F_{\mathrm v\alpha},
\qquad
\mathbbm B
=
\mathbbm B_{\mathrm e\alpha}
+
\mathbbm B_{\mathrm v\alpha},
\qquad
\alpha=1,\ldots,N .
\end{equation}
The corresponding internal variables are
\begin{equation}\label{int_var}
\boldsymbol C_{\mathrm v\alpha}
=
\boldsymbol F_{\mathrm v\alpha}^{T}
\boldsymbol F_{\mathrm v\alpha},
\qquad
\mathbbm B_{\mathrm e\alpha}
=
\mathbbm B-\mathbbm B_{\mathrm v\alpha},
\end{equation}
where \(\boldsymbol C_{\mathrm v\alpha}\) governs mechanical relaxation and \(\mathbbm B_{\mathrm v\alpha}\) governs magnetic relaxation.

The total stored energy is written as
\begin{equation}
\Omega
=
\Omega_{\infty}(\boldsymbol C,\mathbbm B)
+
\sum_{\alpha=1}^{N}
\Omega_{\alpha}
(\boldsymbol C,\boldsymbol C_{\mathrm v\alpha},\mathbbm B,\mathbbm B_{\mathrm v\alpha}),
\end{equation}
where \(\Omega_{\infty}\) is the equilibrium elastic--magnetostatic contribution and \(\Omega_{\alpha}\) is the non-equilibrium contribution of the \(\alpha\)-th branch.

Using the present review notation, a representative equilibrium energy may be written as
\begin{equation}
\Omega_{\infty}
=
\frac{\mu_{\infty}}{4}
\left[
1+\eta_{\infty}
\tanh\left(\frac{I_4}{m_{\infty}}\right)
\right]
\left[
(1+n)(I_1-3)
+
(1-n)(I_2-3)
\right]
+
\frac{a_{\infty}}{2} I_4
+
\frac{b_{\infty}}{2} I_6 .
\end{equation}
Here, \(n\) is the Mooney--Rivlin parameter, while \(\eta_{\infty}\) and \(m_{\infty}\) control field-induced stiffening and saturation-type behaviour. The relevant magnetic invariants are
\begin{equation}\label{B_mag_invar}
I_4
=
\mathbbm B\cdot\mathbbm B,
\qquad
I_6
=
(\boldsymbol C\mathbbm B)\cdot(\boldsymbol C\mathbbm B).
\end{equation}

For each non-equilibrium branch, a representative neo-Hookean-type magneto-viscoelastic energy is
\begin{equation}
\Omega_{\alpha}
=
\frac{\mu_{\alpha}}{2}
\left[
\boldsymbol C_{\mathrm v\alpha}^{-1}:\boldsymbol C
-
3
\right]
+
\frac{a_{\alpha}}{2}
\mathbbm B_{\mathrm e\alpha}\cdot\mathbbm B_{\mathrm e\alpha}
+
\frac{b_{\alpha}}{2}
(\boldsymbol C\mathbbm B_{\mathrm e\alpha})
\cdot
(\boldsymbol C\mathbbm B_{\mathrm e\alpha}) .
\end{equation}
Equivalently, using \(\mathbbm B_{\mathrm e\alpha}=\mathbbm B-\mathbbm B_{\mathrm v\alpha}\),
\begin{equation}
\Omega_{\alpha}
=
\frac{\mu_{\alpha}}{2}
\left[
\boldsymbol C_{\mathrm v\alpha}^{-1}:\boldsymbol C
-
3
\right]
+
\frac{a_{\alpha}}{2}
(\mathbbm B-\mathbbm B_{\mathrm v\alpha})
\cdot
(\mathbbm B-\mathbbm B_{\mathrm v\alpha})
+
\frac{b_{\alpha}}{2}
\left[
\boldsymbol C(\mathbbm B-\mathbbm B_{\mathrm v\alpha})
\right]
\cdot
\left[
\boldsymbol C(\mathbbm B-\mathbbm B_{\mathrm v\alpha})
\right].
\end{equation}

Thermodynamic admissibility requires
\begin{equation}
\mathcal D
=
-
\sum_{\alpha=1}^{N}
\frac{\partial \Omega}{\partial \boldsymbol C_{\mathrm v\alpha}}
:
\dot{\boldsymbol C}_{\mathrm v\alpha}
-
\sum_{\alpha=1}^{N}
\frac{\partial \Omega}{\partial \mathbbm B_{\mathrm v\alpha}}
\cdot
\dot{\mathbbm B}_{\mathrm v\alpha}
\geq 0 .
\end{equation}
A standard magnetic relaxation law satisfying this inequality is
\begin{equation}
\dot{\mathbbm B}_{\mathrm v\alpha}
=
-
\frac{\mu_0}{T_{\mathrm m\alpha}}
\frac{\partial \Omega_{\alpha}}{\partial \mathbbm B_{\mathrm v\alpha}},
\end{equation}
which gives
\begin{equation}
\dot{\mathbbm B}_{\mathrm v\alpha}
=
\frac{\mu_0}{T_{\mathrm m\alpha}}
\left[
a_{\alpha}\boldsymbol I
+
b_{\alpha}\boldsymbol C^2
\right]
(\mathbbm B-\mathbbm B_{\mathrm v\alpha}) .
\end{equation}
The mechanical internal variable may be evolved through
\begin{equation}
\dot{\boldsymbol C}_{\mathrm v\alpha}
=
\frac{1}{T_{\alpha}}
\left[
\boldsymbol C
-
\frac{1}{3}
\left(
\boldsymbol C:\boldsymbol C_{\mathrm v\alpha}^{-1}
\right)
\boldsymbol C_{\mathrm v\alpha}
\right],
\end{equation}
where \(T_{\alpha}\) and \(T_{\mathrm m\alpha}\) are the mechanical and magnetic relaxation times of the \(\alpha\)-th branch.

The constitutive equations are then obtained from the total energy as given by equations \eqref{incomp_const}
and \eqref{const_rel_mag},
where the stress and magnetic field therefore contain both equilibrium and non-equilibrium contributions.

\paragraph{\textbf{Haldar model}}
Haldar and co-workers developed a finite-strain magneto-viscoelastic model for magneto-active polymers with particular emphasis on field-dependent stiffness, damping, demagnetization correction and field-induced Poynting effects \cite{haldarKieferMenzel2016,haldar2021}. In contrast to induction-based formulations, the model is written in terms of the material magnetic field \(\mathbbm H\). The internal variables are the viscous right Cauchy--Green tensors \(\boldsymbol C_{\mathrm v\alpha}\), and the magnetic coupling enters both the equilibrium and non-equilibrium parts of the free energy.

For each dissipative branch, the deformation gradient can be written as given by equation \eqref{def_grad_vis_decomp}.
The deformation gradient is decomposed into volumetric and isochoric parts as $\bar{\boldsymbol F}=J^{-1/3}\boldsymbol F$ and thereby, $\bar{\boldsymbol C}
=
J^{-2/3}\boldsymbol C$.
The viscous part is assumed to be isochoric:
\begin{equation}
\bar{\boldsymbol F}
=
\bar{\boldsymbol F}_{\mathrm e\alpha}
\boldsymbol F_{\mathrm v\alpha},
\qquad
\det\bar{\boldsymbol F}_{\mathrm e\alpha}
=
\det\boldsymbol F_{\mathrm v\alpha}
=
1,
\end{equation}
and the internal variable $\boldsymbol C_{\mathrm v\alpha}$ is given equation \eqref{int_var}.

The Helmholtz free energy per unit reference volume is written as
\begin{equation}
\Omega
=
\Omega_{\mathrm{vol}}(J)
+
\Omega_{\infty}^{\mathrm E}(\bar{\boldsymbol C})
+
\Omega_{\infty}^{\mathrm M}(\bar{\boldsymbol C},\mathbbm H)
+
\Omega^{\mathrm H}(\mathbbm H)
+
\sum_{\alpha=1}^{N}
\left[
\Omega_{\alpha}^{\mathrm E}(\bar{\boldsymbol C},\boldsymbol C_{\mathrm v\alpha})
+
\Omega_{\alpha}^{\mathrm M}(\bar{\boldsymbol C},\boldsymbol C_{\mathrm v\alpha},\mathbbm H)
\right].
\end{equation}
The terms denote volumetric, equilibrium elastic, equilibrium magneto-elastic, purely magnetic, non-equilibrium elastic and non-equilibrium magneto-viscoelastic contributions, respectively.

The isotropic invariants are
\begin{equation}
\begin{aligned}
I_1 &= \mathrm{tr}(\bar{\boldsymbol C}), \qquad
&I_2 &= \frac{1}{2}
\left[
I_1^2-\mathrm{tr}(\bar{\boldsymbol C}^{2})
\right],\\
I_4 &= \mathbbm H\cdot\mathbbm H, \qquad
&I_5 &= \mathbbm H\cdot\bar{\boldsymbol C}\mathbbm H, \qquad
&I_6 &= \mathbbm H\cdot\bar{\boldsymbol C}^{2}\mathbbm H .
\end{aligned}
\end{equation}
For the saturation-type magnetostrictive term, the model also uses
\begin{equation}
\bar I_5
=
\mathbbm H\cdot\bar{\boldsymbol C}^{-1}\mathbbm H .
\end{equation}

A representative volumetric penalty may be written as
\begin{equation}
\Omega_{\mathrm{vol}}
=
\frac{\kappa}{2}(J-1)^2.
\end{equation}
 The equilibrium elastic part is chosen in Mooney--Rivlin form:
\begin{equation}
\Omega_{\infty}^{\mathrm E}
=
C_1(I_1-3)
+
C_2(I_2-3),
\end{equation}
with $C_1$ and $C_2$ as material constants.
The equilibrium magneto-elastic contribution is
\begin{equation}
\Omega_{\infty}^{\mathrm M}
=
\widehat C_2(I_4)(I_2-3)
-
\frac{\mu_0}{2}
\left[
\frac{\gamma}{\beta}
\ln(1+\beta\bar I_5)
-
\bar I_5
\right],
\end{equation}
with
\begin{equation}
\widehat C_2(I_4)
=
\zeta_2
\arctan
\left(
\eta_2\sqrt{I_4}
\right).
\end{equation}
The first term introduces field-dependent stiffening, while the second represents saturation-type magnetostrictive coupling.

The purely magnetic contribution is
\begin{equation}
\Omega^{\mathrm H}
=
-\frac{\mu_0}{2}I_4
-
\frac{\nu_1}{k_1k_2}
\sqrt{I_4}
\arctan
\left(
\frac{k_2}{k_1}\sqrt{I_4}
\right)
+
\frac{\nu_1}{2k_2^2}
\ln
\left(
1+
\frac{k_2^2}{k_1^2}I_4
\right).
\end{equation}

For the \(\alpha\)-th branch, the non-equilibrium mechanical contribution is
\begin{equation}
\Omega_{\alpha}^{\mathrm E}
=
\frac{\mu_{\mathrm v\alpha}}{2}
\left[
\bar{\boldsymbol C}:\boldsymbol C_{\mathrm v\alpha}^{-1}
-
3
\right],
\end{equation}
and the corresponding magneto-viscoelastic contribution is
\begin{equation}
\Omega_{\alpha}^{\mathrm M}
=
\frac{\zeta_{\mathrm v\alpha}}{2}
\arctan
\left(
\eta_{\mathrm v\alpha}\sqrt{I_4}
\right)
\left[
\bar{\boldsymbol C}:\boldsymbol C_{\mathrm v\alpha}^{-1}
-
3
\right].
\end{equation}
Thus, the branch stiffness is made field-dependent through the factor
\begin{equation}
\mu_{\mathrm v\alpha}
+
\zeta_{\mathrm v\alpha}
\arctan
\left(
\eta_{\mathrm v\alpha}\sqrt{I_4}
\right).
\end{equation}

The evolution of \(\boldsymbol C_{\mathrm v\alpha}\) follows from
\begin{equation}
\mathcal D
=
-
\sum_{\alpha=1}^{N}
\frac{\partial \Omega}{\partial \boldsymbol C_{\mathrm v\alpha}}
:
\dot{\boldsymbol C}_{\mathrm v\alpha}
\geq 0 .
\end{equation}
Introducing the Mandel-type driving stress
\begin{equation}
\boldsymbol\Sigma_{\mathrm v\alpha}
=
-
\boldsymbol C_{\mathrm v\alpha}
\frac{\partial \Omega}{\partial \boldsymbol C_{\mathrm v\alpha}},
\end{equation}
a thermodynamically admissible evolution law is
\begin{equation}
\dot{\boldsymbol C}_{\mathrm v\alpha}
=
\dot{\Gamma}_{\alpha}
\boldsymbol C_{\mathrm v\alpha}
\boldsymbol\Sigma_{\mathrm v\alpha}^{\mathrm{dev}\,T},
\qquad
\dot{\Gamma}_{\alpha}>0 .
\end{equation}
For the energy above, this may be written in the form
\begin{equation}
\dot{\boldsymbol C}_{\mathrm v\alpha}
=
\frac{\dot{\Gamma}_{\alpha}}{2}
\left[
\mu_{\mathrm v\alpha}
+
\zeta_{\mathrm v\alpha}
\arctan
\left(
\eta_{\mathrm v\alpha}\sqrt{I_4}
\right)
\right]
\left[
\bar{\boldsymbol C}
-
\frac{1}{3}
\left(
\bar{\boldsymbol C}:\boldsymbol C_{\mathrm v\alpha}^{-1}
\right)
\boldsymbol C_{\mathrm v\alpha}
\right].
\end{equation}
The constitutive equations are then obtained from the total energy as given by equations \eqref{const_rel_mag} and \eqref{incomp_const}.
The magnetization follows from
\begin{equation}
\mathbbm M
=
\frac{1}{\mu_0}\mathbbm B-\mathbbm H,
\end{equation}
 containing contributions from the equilibrium magneto-elastic, non-equilibrium magneto-viscoelastic and purely magnetic energy terms.

For comparison with experiments, the field entering the constitutive equations is interpreted as the corrected internal magnetic field. The demagnetization correction may be written compactly as
\begin{equation}
\mathbbm H_{\mathrm{int}}
=
\mathbbm H_{\mathrm{app}}
-
\boldsymbol N_{\mathrm d}\mathbbm M,
\end{equation}
where \(\mathbbm H_{\mathrm{app}}\) is the externally applied field, \(\boldsymbol N_{\mathrm d}\) is the demagnetization tensor and \(\mathbbm M\) is the magnetization. The same constitutive structure was later used to analyse field-induced Poynting effects in simple shear, including the possibility of a negative Poynting response.

\paragraph{\textbf{Nedjar model}}
Nedjar proposed a thermodynamically consistent framework for finite-strain magneto-viscoelasticity of isotropic magnetizable soft materials \cite{nedjar2020}. The model is based on the magnetic induction as the independent magnetic variable and uses the multiplicative decomposition of the deformation gradient to define an intermediate configuration. Unlike formulations that introduce an additive split of the magnetic induction, Nedjar transports the induction to the intermediate configuration. The corresponding non-equilibrium magnetic-field contribution is then induced through this transport.

The kinematics are:
\begin{equation}\label{nejdar_1}
\boldsymbol F
=
\boldsymbol F_{\mathrm e}
\boldsymbol F_{\mathrm v},
\qquad
\boldsymbol C_{\mathrm e}
=
\boldsymbol F_{\mathrm e}^{T}\boldsymbol F_{\mathrm e}
=
\boldsymbol F_{\mathrm v}^{-T}
\boldsymbol C
\boldsymbol F_{\mathrm v}^{-1}.
\end{equation}
The magnetic induction \(\mathbbm B\), defined in the reference configuration, is transported to the intermediate configuration as
\begin{equation}
\widetilde{\mathbbm B}
=
J_{\mathrm v}^{-1}
\boldsymbol F_{\mathrm v}\mathbbm B,
\qquad
J_{\mathrm v}
=
\det\boldsymbol F_{\mathrm v}.
\end{equation}

The augmented free energy is written as
\begin{equation}
\Omega
=
\Omega_{\infty}
+
\Omega_{\mathrm v},
\end{equation}
with
\begin{equation}
\Omega_{\infty}
=
\Omega_{\infty}(\boldsymbol C,\mathbbm B),
\qquad
\Omega_{\mathrm v}
=
\Omega_{\mathrm v}(\boldsymbol C_{\mathrm e},\widetilde{\mathbbm B}).
\end{equation}
No branch index is introduced in this model because it presents a single non-equilibrium contribution.
For the equilibrium part, $I_1$, $I_2$ and $I_3$  are given from the equation set \eqref{definvar}, and $I_4$, $I_6$ are defined by equation \eqref{B_mag_invar}. $I_5$ is defined as $I_5 = \mathbbm B\cdot(\boldsymbol C\mathbbm B)$
The corresponding intermediate-configuration invariants are:
\begin{equation}
\begin{aligned}
\widetilde I_1 &= \mathrm{tr}(\boldsymbol C_{\mathrm e}), \qquad
&\widetilde I_2 &= \frac{1}{2}
\left[
\widetilde I_1^2-\mathrm{tr}(\boldsymbol C_{\mathrm e}^{2})
\right], \qquad
&\widetilde I_3 &= \det\boldsymbol C_{\mathrm e},\\
\widetilde I_4 &= \widetilde{\mathbbm B}\cdot\widetilde{\mathbbm B}, \qquad
&\widetilde I_5 &= \widetilde{\mathbbm B}\cdot(\boldsymbol C_{\mathrm e}\widetilde{\mathbbm B}), \qquad
&\widetilde I_6 &= \widetilde{\mathbbm B}\cdot(\boldsymbol C_{\mathrm e}^{2}\widetilde{\mathbbm B}).
\end{aligned}
\end{equation}

A representative energy form is
\begin{equation}
\Omega_{\infty}
=
\Omega_{\infty}^{\mathrm{mech}}(I_1,I_2,I_3)
+
c_1 I_4
+
c_2 I_5
+
c_3 I_6
+
\frac{1}{2\mu_0}
J^{-1}
\boldsymbol C:(\mathbbm B\otimes\mathbbm B),
\end{equation}
and
\begin{equation}
\Omega_{\mathrm v}
=
\Omega_{\mathrm v}^{\mathrm{mech}}(\widetilde I_1,\widetilde I_2,\widetilde I_3)
+
c_4 \widetilde I_4
+
c_5 \widetilde I_5
+
c_6 \widetilde I_6 .
\end{equation}
Here, \(c_1,c_2,c_3\) are equilibrium magnetic coupling parameters, while \(c_4,c_5,c_6\) are the corresponding non-equilibrium parameters. The final term in \(\Omega_{\infty}\) is the augmentation term associated with the magnetic-induction formulation.

For the demonstrative model, Nedjar used compressible neo-Hookean-type mechanical energies:
\begin{equation}
\Omega_{\infty}^{\mathrm{mech}}
=
\frac{3}{8}\kappa_{\infty}
\left[
J^{4/3}
+
2J^{-2/3}
-
3
\right]
+
\frac{1}{2}\mu_{\infty}
\left[
J^{-2/3}I_1
-
3
\right],
\end{equation}
and
\begin{equation}
\Omega_{\mathrm v}^{\mathrm{mech}}
=
\frac{3}{8}\kappa_{\mathrm v}
\left[
J_{\mathrm e}^{4/3}
+
2J_{\mathrm e}^{-2/3}
-
3
\right]
+
\frac{1}{2}\mu_{\mathrm v}
\left[
J_{\mathrm e}^{-2/3}\widetilde I_1
-
3
\right],
\end{equation}
where \(J_{\mathrm e}=\det\boldsymbol F_{\mathrm e}\).

The reduced dissipation inequality governs the evolution of the viscous internal variable. With
\begin{equation}
\boldsymbol b_{\mathrm e}
=
\boldsymbol F_{\mathrm e}\boldsymbol F_{\mathrm e}^{T},
\qquad
\boldsymbol C_{\mathrm v}
=
\boldsymbol F_{\mathrm v}^{T}\boldsymbol F_{\mathrm v},
\end{equation}
and denoting the spatial non-equilibrium magnetic field by \(\mathbbm h_{\mathrm v}\), Nedjar writes the dissipation in the form
\begin{equation}
\mathcal D
=
\left[
\boldsymbol\tau_{\mathrm v}
-
J\,\mathbbm h_{\mathrm v}\otimes\mathbbm b
+
J(\mathbbm h_{\mathrm v}\cdot\mathbbm b)\boldsymbol i
\right]
\boldsymbol b_{\mathrm e}^{-1}
:
\left(
-\frac{1}{2}
\mathcal L_{\mathrm v}\boldsymbol b_{\mathrm e}
\right)
\geq 0 .
\end{equation}
A thermodynamically admissible evolution law is obtained by choosing a positive-definite fourth-order mobility tensor \(\mathbb M\):
\begin{equation}
-\frac{1}{2}
\mathcal L_{\mathrm v}\boldsymbol b_{\mathrm e}
=
\mathbb M
:
\left\{
\left[
\boldsymbol\tau_{\mathrm v}
-
J\,\mathbbm h_{\mathrm v}\otimes\mathbbm b
+
J(\mathbbm h_{\mathrm v}\cdot\mathbbm b)\boldsymbol i
\right]
\boldsymbol b_{\mathrm e}^{-1}
\right\}.
\end{equation}
For the simple demonstrative case,
\begin{equation}
\mathbb M
=
\frac{1}{\eta}\mathbb I,
\end{equation}
where \(\eta>0\) is a viscosity parameter. The evolution law then becomes
\begin{equation}
-\frac{1}{2}
\mathcal L_{\mathrm v}\boldsymbol b_{\mathrm e}
=
\frac{1}{\eta}
\left[
\boldsymbol\tau_{\mathrm v}^{\mathrm{mech}}\boldsymbol b_{\mathrm e}^{-1}
-
2c_4J_{\mathrm e}^{2}
\boldsymbol b_{\mathrm e}^{-1}
\mathbbm b\otimes\mathbbm b
\boldsymbol b_{\mathrm e}^{-1}
+
2c_6J_{\mathrm e}^{2}
\mathbbm b\otimes\mathbbm b
+
2
\left(
c_4\widetilde I_4
+
c_5\widetilde I_5
+
c_6\widetilde I_6
\right)
\boldsymbol b_{\mathrm e}^{-1}
\right].
\end{equation}

The constitutive equations are obtained by differentiating the augmented energy.
Equivalently, the second Piola--Kirchhoff-type stress is
\begin{equation}
\boldsymbol S^{\mathrm{tot}}
=
2\frac{\partial\Omega_{\infty}}{\partial\boldsymbol C}
+
\boldsymbol F_{\mathrm v}^{-1}
\left[
2\frac{\partial\Omega_{\mathrm v}}{\partial\boldsymbol C_{\mathrm e}}
\right]
\boldsymbol F_{\mathrm v}^{-T},
\end{equation}
with \(-p\boldsymbol C^{-1}\) added for incompressibility. The material magnetic field is
\begin{equation}
\mathbbm H
=
\frac{\partial\Omega_{\infty}}{\partial\mathbbm B}
+
J_{\mathrm v}^{-1}
\boldsymbol F_{\mathrm v}^{T}
\frac{\partial\Omega_{\mathrm v}}{\partial\widetilde{\mathbbm B}} .
\end{equation}
This transport-induced non-equilibrium magnetic term is the key distinction between Nedjar's formulation and models based on an explicit additive decomposition of \(\mathbbm B\).

\paragraph{\textbf{Garcia-Gonzalez--Hossain microstructural magneto-viscoelastic model}}
Garcia-Gonzalez and Hossain developed a microstructurally based finite-strain magneto-viscoelastic model for magneto-active polymers \cite{garciaGonzalezHossain2021}. The model links the macroscopic response to the arrangement of magnetizable particles in the polymeric matrix. The matrix provides the elastic and viscous mechanical response, while the magnetic contribution is obtained from dipole--dipole interactions between particles and transferred to the continuum scale through an affine deformation assumption.

The total energy per unit reference volume is written as
\begin{equation}
\Omega
=
\Omega_{\infty}
+
\Omega_{\mathrm v},
\end{equation}
with
\begin{equation}
\Omega_{\infty}
=
\Omega_{\mathrm m}^{\mathrm e}(\boldsymbol F)
+
\Omega_{\mathrm{mg}}(\boldsymbol F,\mathbbm M),
\qquad
\Omega_{\mathrm v}
=
\Omega_{\mathrm m}^{\mathrm v}(\boldsymbol F,\boldsymbol F_{\mathrm v}).
\end{equation}
Thus,
\begin{equation}
\Omega
=
\Omega_{\mathrm m}^{\mathrm e}(\boldsymbol F)
+
\Omega_{\mathrm m}^{\mathrm v}(\boldsymbol F,\boldsymbol F_{\mathrm v})
+
\Omega_{\mathrm{mg}}(\boldsymbol F,\mathbbm M),
\end{equation}
where \(\mathbbm M\) is the material magnetization.

The finite-strain viscoelastic response is described using the same kinematic framework as in the Nedjar model.
Let \(\phi\) be the volume fraction of magnetizable particles, so that \(1-\phi\) is the polymeric volume fraction. The mechanical energy is scaled by \(1-\phi\), with an additional amplification factor in the elastic part to account for particle-induced stiffening.

For the Arruda--Boyce 8-chain representation,
\begin{equation}
\Omega_{\mathrm m,8c}^{\mathrm e}
=
(1-\phi)G^{\mathrm e}
\sum_{k=1}^{K}
\frac{C_k}{N_{\mathrm e}^{k-1}}
\left[
I_{1,h}^{k}
-
3^{k}
\right],
\end{equation}
and
\begin{equation}
\Omega_{\mathrm m,8c}^{\mathrm v}
=
(1-\phi)G^{\mathrm v}
\sum_{k=1}^{K}
\frac{C_k}{N_{\mathrm v}^{k-1}}
\left[
(I_1^{\mathrm e})^{k}
-
3^{k}
\right].
\end{equation}
Here,
\begin{equation}
I_1^{\mathrm e}
=
\mathrm{tr}(\boldsymbol C_{\mathrm e}),
\qquad
I_{1,h}
=
X(I_1-3)+3,
\qquad
I_1=\mathrm{tr}(\boldsymbol C),
\end{equation}
with
\begin{equation}
X
=
1+0.67\,g\phi+1.62(g\phi)^2 .
\end{equation}
The constants \(G^{\mathrm e}\), \(G^{\mathrm v}\), \(N_{\mathrm e}\), \(N_{\mathrm v}\) and \(C_k\) are the elastic and viscous chain-network parameters, while \(g\) accounts for asymmetric particle aggregates.


The equilibrium magnetic energy is constructed from dipole--dipole interactions. At the microscale,
\begin{equation}
\Omega_{\mathrm{mg}}
=
-\frac{1}{V_0}
\frac{\mu_r\mu_0}{4\pi}
\sum_{i,j}
\left[
\frac{
3(\boldsymbol d_i\cdot\boldsymbol r_{ij})
(\boldsymbol d_j\cdot\boldsymbol r_{ij})
}
{\|\boldsymbol r_{ij}\|^5}
-
\frac{
\boldsymbol d_i\cdot\boldsymbol d_j
}
{\|\boldsymbol r_{ij}\|^3}
\right],
\end{equation}
where \(V_0\) is the reference volume, \(\boldsymbol d_i\) and \(\boldsymbol d_j\) are dipole moments and \(\boldsymbol r_{ij}\) is the deformed separation vector between particles. Under the affine assumption,
\begin{equation}
\boldsymbol r_i
=
\boldsymbol F\boldsymbol R_i.
\end{equation}
For an idealized infinite lattice with identical particle magnetization,
\begin{equation}
\Omega_{\mathrm{mg}}(\boldsymbol F,\mathbbm M)
=
-\frac{\mu_r\mu_0}{4\pi}
\frac{\phi^2}{\gamma}
\sum_{i=1}^{N}
\left[
\frac{
3\left[
(\boldsymbol F^{-T}\mathbbm M)\cdot
(\boldsymbol F\boldsymbol R_i^0)
\right]^2
}
{\|\boldsymbol F\boldsymbol R_i^0\|^5}
-
\frac{
(\boldsymbol F^{-T}\mathbbm M)\cdot
(\boldsymbol F^{-T}\mathbbm M)
}
{\|\boldsymbol F\boldsymbol R_i^0\|^3}
\right].
\end{equation}
The reference vectors \(\boldsymbol R_i^0\) encode the particle morphology, allowing the same framework to describe isotropic distributions, perfectly aligned chains and wavy chains.

The non-equilibrium evolution is governed by a flow rule satisfying the dissipation inequality associated with \(\boldsymbol F_{\mathrm v}\). In compact form,
\begin{equation}
\boldsymbol D_{\mathrm v}
=
\frac{1}{\sqrt{2}\eta}
\boldsymbol\sigma_{\mathrm m}^{\mathrm v},
\end{equation}
where \(\eta\) is the polymer viscosity and \(\boldsymbol\sigma_{\mathrm m}^{\mathrm v}\) is the viscous Cauchy stress contribution.

Using the decomposition of \(\Omega\), the total first Piola--Kirchhoff stress may be written as
\begin{equation}
\boldsymbol P^{\mathrm{tot}}
=
\boldsymbol P_{\mathrm m}^{\mathrm e}
+
\boldsymbol P_{\mathrm{mg}}
+
\boldsymbol P_{\mathrm m}^{\mathrm v}
-
p\boldsymbol F^{-T},
\end{equation}
where
\begin{equation}
\boldsymbol P_{\mathrm m}^{\mathrm e}
=
\frac{\partial \Omega_{\mathrm m}^{\mathrm e}}{\partial \boldsymbol F},
\qquad
\boldsymbol P_{\mathrm{mg}}
=
\frac{\partial \Omega_{\mathrm{mg}}}{\partial \boldsymbol F},
\qquad
\boldsymbol P_{\mathrm m}^{\mathrm v}
=
\frac{\partial \Omega_{\mathrm m}^{\mathrm v}}{\partial \boldsymbol F_{\mathrm e}}
\boldsymbol F_{\mathrm v}^{-T}.
\end{equation}
The model is naturally formulated in terms of magnetization. Garcia-Gonzalez and Hossain use the Froehlich--Kennelly relation,
\begin{equation}
\mathbbm M
=
M_s
\frac{(\mu_r-1)\mathbbm H}
{M_s+(\mu_r-1)\|\mathbbm H\|},
\end{equation}
where \(M_s\) is the saturation magnetization and \(\mu_r\) is the relative magnetic permeability. This relation introduces magnetic saturation into the model by linking the material magnetization to the material magnetic field. The main value of the formulation is that the particle arrangement enters directly through \(\boldsymbol R_i^0\), thereby connecting finite-strain viscoelasticity, magnetic saturation and microstructural dipole interactions.

\subsubsection{Anisotropic elastic constitutive models}\label{constvmodelaniso}


\paragraph{\textbf{Bustamante model}}
The extension from isotropic to anisotropic constitutive modelling is essential for magneto-active polymers cured in the presence of an applied magnetic field. During curing, magnetizable particles may align along the field direction and form chain-like microstructures, thereby introducing a preferred material direction and a transversely isotropic magneto-mechanical response. Bustamante developed one of the early finite-strain constitutive frameworks for such materials by extending the nonlinear magnetoelastic theory of Dorfmann and Ogden to include the preferred particle-chain direction \cite{bustamante1}. In the notation used here, \(\boldsymbol a_0\) denotes the unit preferred direction in the reference configuration.

Using the material magnetic field \(\mathbbm H\) as the independent magnetic variable, the amended energy may be written as
\begin{equation}
\Omega
=
\Omega^{*}(\boldsymbol F,\mathbbm H,\boldsymbol a_0).
\end{equation}
Equivalently, \(\Omega^{*}\) may be expressed in terms of invariants of the right Cauchy--Green tensor
\(\boldsymbol C\), the material magnetic field \(\mathbbm H\), and the preferred direction \(\boldsymbol a_0\). A representative invariant set is
\begin{equation}
\begin{aligned}
I_1 &= \mathrm{tr}(\boldsymbol C), \qquad
&I_2 &= \frac{1}{2}\left[I_1^2-\mathrm{tr}(\boldsymbol C^2)\right], \qquad
&I_3 &= \det(\boldsymbol C),\\
I_4 &= \mathbbm H\cdot\mathbbm H, \qquad
&I_5 &= \mathbbm H\cdot(\boldsymbol C\mathbbm H), \qquad
&I_6 &= \mathbbm H\cdot(\boldsymbol C^2\mathbbm H),\\
I_7 &= \boldsymbol a_0\cdot(\boldsymbol C\boldsymbol a_0), \qquad
&I_8 &= \boldsymbol a_0\cdot(\boldsymbol C^2\boldsymbol a_0), \qquad
&I_9 &= \boldsymbol a_0\cdot\mathbbm H,\\
I_{10} &= \boldsymbol a_0\cdot(\boldsymbol C\mathbbm H).
\end{aligned}
\end{equation}
Here, \(I_1\), \(I_2\), and \(I_3\) are mechanical invariants, \(I_4\), \(I_5\), and \(I_6\) describe the magnetic and isotropic magneto-mechanical response, while \(I_7\), \(I_8\), \(I_9\), and \(I_{10}\) introduce the preferred-direction dependence. Thus,
\begin{equation}
\Omega^{*}
=
\Omega^{*}
(I_1,I_2,I_3,I_4,I_5,I_6,I_7,I_8,I_9,I_{10}).
\end{equation}
For incompressible materials, and because the full invariant dependence is difficult to identify from limited experimental data, Bustamante proposed reduced forms involving the dominant mechanical, magnetic, and anisotropic coupling invariants. A reduced energy may be written as
\begin{equation}
\Omega^{*}
=
\Omega^{*}(I_1,I_4,I_5,I_7,I_9,I_{10}).
\end{equation}
Following the usual additive separation of isotropic and anisotropic contributions, this may be decomposed as
\begin{equation}
\Omega^{*}
=
\widehat{\Omega}^{*}(I_1,I_4,I_5)
+
\widetilde{\Omega}^{*}(I_7,I_9,I_{10}),
\end{equation}
where
\begin{equation}
\widehat{\Omega}^{*}(I_1,I_4,I_5)
=
f(I_1)g(I_4)
+
\nu(I_4)
+
\vartheta(I_5),
\end{equation}
and
\begin{equation}
\widetilde{\Omega}^{*}(I_7,I_9,I_{10})
=
h(I_7)\omega(I_9,I_{10})
+
\eta(I_9).
\end{equation}
The function \(f(I_1)\) controls the isotropic elastic response, \(g(I_4)\) introduces field-dependent stiffening, \(\nu(I_4)\) gives the purely magnetic part, and \(\vartheta(I_5)\) introduces isotropic magneto-mechanical coupling. The anisotropic part is governed by \(h(I_7)\), which captures deformation along the preferred direction, \(\omega(I_9,I_{10})\), which couples the preferred direction to the magnetic field and deformation, and \(\eta(I_9)\), which accounts for magnetic alignment with the chain direction.


A prototype choice used by Bustamante may be represented by
\begin{equation}
f(I_1)
=
\frac{1}{k}
\left[
\frac{(I_1-3)^k}{2k-1}
\right],
\qquad
g(I_4)=g_0+g_1I_4,
\qquad
\vartheta(I_5)=\frac{\mu_0}{2}I_5,
\end{equation}
and
\begin{equation}
h(I_7)
=
h_0+h_1\ln I_7-\frac{h_1}{m}I_7^m,
\qquad
m<0.
\end{equation}
The coupling function may be approximated by
\begin{equation}
\omega(I_9,I_{10})
=
\omega_0+\omega_1I_9^2+\omega_2I_{10}^2+\omega_3I_9I_{10}.
\end{equation}
A simple admissible choice for the alignment energy is
\begin{equation}
\eta(I_9)
=
\eta_0 I_9^2,
\end{equation}
with \(\eta_0\) a material parameter.

For incompressible materials, the constitutive equations follow from the amended energy as given by equations \eqref{const_rel_mag} and \eqref{incomp_const}.
The later work of Salas and Bustamante \cite{salasBustamante2015} should be viewed as a numerical boundary-value application of this modelling framework.

\paragraph{\textbf{Danas--Kankanala--Triantafyllidis model}}
Danas, Kankanala and Triantafyllidis proposed an experimentally calibrated finite-strain model for transversely isotropic magnetoactive elastomers cured in the presence of a strong magnetic field \cite{danas1}. The curing field produces chain-like particle structures, and the resulting material response depends on the deformation, the magnetization, and the preferred chain direction. In the present notation, the energy per unit reference volume is written as
\begin{equation}
\Omega
=
\Omega(\boldsymbol C,\mathbbm M,\boldsymbol A),
\end{equation}
where \(\mathbbm M\) is the material magnetization and \(\boldsymbol A\) is the unit chain direction in the reference configuration.

For nearly incompressible response, the model may be expressed through a reduced set of transversely isotropic invariants:
\begin{equation}
\Omega
=
\mathcal C(I_1,I_4,I_6,I_7,I_8,I_9,I_{10}),
\end{equation}
where
\begin{equation}
\begin{aligned}
I_1 &= \mathrm{tr}(\boldsymbol C), \qquad
&I_4 &= \boldsymbol A\cdot(\boldsymbol C\boldsymbol A),\\
I_6 &= \mathbbm M\cdot\mathbbm M, \qquad
&I_7 &= \mathbbm M\cdot(\boldsymbol C\mathbbm M), \qquad
&I_8 &= \mathbbm M\cdot(\boldsymbol C^2\mathbbm M),\\
I_9 &= (\mathbbm M\cdot\boldsymbol A)^2, \qquad
&I_{10} &=
(\mathbbm M\cdot\boldsymbol A)
\left[
\mathbbm M\cdot(\boldsymbol C\boldsymbol A)
\right].
\end{aligned}
\end{equation}
The invariant \(I_4\) captures the stretch of the particle-chain direction, \(I_6\) is the purely magnetic invariant, and \(I_7\), \(I_8\), \(I_9\), and \(I_{10}\) encode magneto-mechanical coupling and directional anisotropy.

The energy is constructed to reproduce the experimentally measured magnetization, magnetostriction, and shear responses for different particle-chain orientations and magnetic-field directions. A compact way to express the structure of the model is
\begin{equation}
\Omega
=
\Omega_{\mathrm{mech}}(I_1,I_4)
+
\Omega_{\mathrm{mag}}(I_6)
+
\Omega_{\mathrm{coup}}(I_7,I_8,I_9,I_{10}),
\end{equation}
where the first term represents the transversely isotropic mechanical response, the second captures magnetic storage and saturation, and the third introduces the deformation-dependent magneto-mechanical coupling.

The constitutive equations follow from the energy by differentiation with respect to the deformation and magnetization variables. For an incompressible material, $\boldsymbol S^{\mathrm{tot}}$ is given by equation \eqref{incomp_const},
and the material magnetic field is obtained from
\begin{equation*}
\mathbbm H
=
\frac{\partial \Omega}{\partial \mathbbm M},
\end{equation*}
up to the conventional permeability or density factors used in the original variational formulation. The main significance of this model is that it is calibrated directly against experiments on anisotropic particle-chain elastomers and can capture magnetization, magnetostriction and shear responses within the same finite-strain constitutive setting.

\paragraph{\textbf{Saxena--Pelteret--Steinmann dispersed-chain elastic model}}
The dispersed-chain model of Saxena, Pelteret and Steinmann was introduced to represent imperfect chain alignment in iron-filled magneto-active polymers \cite{saxena4}. Instead of assuming a perfectly aligned transversely isotropic microstructure, the model allows the chain orientations to be distributed about a preferred mean direction. This dispersion is described through a probability-based generalized structure tensor.

Let \(\boldsymbol M\) denote a unit vector representing a chain orientation. The generalized structure tensor is defined by
\begin{equation}
\boldsymbol G
=
\frac{1}{4\pi}
\int_{\omega}
\chi(\boldsymbol M)\,
\boldsymbol M\otimes\boldsymbol M
\,\mathrm d\omega,
\end{equation}
where \(\chi(\boldsymbol M)\) is the orientation density function over the unit sphere \(\omega\), normalized as
\begin{equation}
\frac{1}{4\pi}
\int_{\omega}
\chi(\boldsymbol M)\,\mathrm d\omega
=
1.
\end{equation}
For a transversely isotropic distribution about the mean chain direction \(\boldsymbol M\), the tensor reduces to
\begin{equation}
\boldsymbol G
=
k_d\boldsymbol I
+
(1-3k_d)\boldsymbol M\otimes\boldsymbol M,
\qquad
0\leq k_d\leq \frac{1}{3}.
\end{equation}
The limiting case \(k_d=0\) corresponds to perfectly aligned chains, while \(k_d=1/3\) corresponds to an isotropic orientation distribution.

The total energy is written as
\begin{equation}
\Omega
=
U_0(\boldsymbol F,\mathbbm H,\boldsymbol G),
\end{equation}
where \(\mathbbm H\) is the material magnetic field. The key modelling step is to decouple the matrix contribution from the chain contribution while keeping both phases kinematically connected:
\begin{equation}
\Omega
=
U_0^{\mathrm{mat}}(\boldsymbol C,\mathbbm H)
+
U_0^{c}(\boldsymbol C^{c},\mathbbm H^{c},\boldsymbol G).
\end{equation}
Here, \(\boldsymbol C^{c}\) and \(\mathbbm H^{c}\) denote the chain-level right Cauchy--Green tensor and chain magnetic field, respectively. They are constructed from the macroscopic fields and the structure tensor so that chain dispersion enters the magnetoelastic response through \(\boldsymbol G\).

A representative form is
\begin{equation}
U_0^{\mathrm{mat}}
=
\frac{\nu_1}{a}
\left[
\exp\left(a(\boldsymbol C:\boldsymbol I-3)^2\right)-1
\right]
+
\nu_2\,
\mathbbm H\otimes\mathbbm H:\boldsymbol I
+
\nu_3\,
\mathbbm H\otimes\mathbbm H:\boldsymbol C^{-1},
\end{equation}
and
\begin{equation}
U_0^{c}
=
\frac{\nu_4}{a}
\left[
\exp\left(a(\boldsymbol C^{c}:\boldsymbol I-\boldsymbol G^2:\boldsymbol I)^2\right)-1
\right]
+
\nu_5\,
\mathbbm H^{c}\otimes\mathbbm H^{c}:\boldsymbol I
+
\nu_6\,
\mathbbm H^{c}\otimes\mathbbm H^{c}:\left(\boldsymbol C^{c}\right)^{-1}.
\end{equation}
The parameters \(\nu_1,\ldots,\nu_6\) and \(a\) control the matrix and chain-level mechanical, magnetic and magneto-mechanical responses. The first term in each energy describes nonlinear elastic storage, while the remaining terms account for magnetic storage and magneto-mechanical coupling.

For an incompressible material, the constitutive equations are obtained from equations \eqref{incomp_const} and \eqref{const_rel_mag}.
The dispersed-chain model therefore provides a physically interpretable route for incorporating imperfect chain alignment, matrix--chain separation and anisotropic magneto-mechanical coupling into a finite-strain elastic framework.

\paragraph{\textbf{Shariff--Bustamante--Hossain--Steinmann spectral model}}
Shariff, Bustamante, Hossain and Steinmann proposed a spectral invariant formulation for transversely isotropic magnetoelasticity \cite{shariff5}. The model is motivated by the fact that classical invariants often lead to long constitutive expressions and do not always provide a direct physical interpretation. The spectral approach instead uses the principal stretches and the projections of the magnetic field and preferred direction onto the principal material axes.

Let the right stretch tensor have principal stretches \(\lambda_a\) and corresponding material principal directions \(\boldsymbol N_a\), with \(a=1,2,3\). The material magnetic field and preferred direction are projected as
\begin{equation}
H_a
=
\mathbbm H\cdot\boldsymbol N_a,
\qquad
A_a
=
\boldsymbol A\cdot\boldsymbol N_a,
\qquad
a=1,2,3,
\end{equation}
where \(\boldsymbol A\) is the unit preferred direction in the reference configuration. The energy may then be written as
\begin{equation}
\Omega
=
\widehat{\Omega}_{\mathrm{Sh}}^{\mathrm{ti}}
(\lambda_1,\lambda_2,\lambda_3,H_1,H_2,H_3,A_1,A_2,A_3).
\end{equation}

A general spectral representation can be expressed in the separable form
\begin{equation}
\Omega
=
\sum_{a=1}^{3}
r(\lambda_a,\zeta_a,\xi_a,\chi_a,\mathbbm H),
\end{equation}
where the spectral invariants \(\zeta_a\), \(\xi_a\), and \(\chi_a\) encode the projections and interactions of the magnetic field and preferred material direction in the principal stretch basis. A representative structure used in the spectral formulation is
\begin{equation}
r(\lambda_a,\zeta_a,\xi_a,\chi_a,\mathbbm H)
=
r_0(\lambda_a)
+
\zeta_a r_1(\lambda_a,\mathbbm H)
+
\xi_a r_2(\lambda_a)
+
\chi_a r_3(\lambda_a,\mathbbm H).
\end{equation}
Here, \(r_0\) represents the isotropic stretch response, \(r_1\) introduces magnetic coupling, \(r_2\) describes the purely anisotropic mechanical contribution, and \(r_3\) captures mixed anisotropic magneto-mechanical coupling. This form reflects the essential feature of the Shariff model: the effects of stretch, magnetic loading and preferred direction can be separated and interpreted componentwise in the principal basis.

For incompressible materials, 
the principal constitutive equations are obtained by differentiating the spectral energy:
\begin{equation}
P_a
=
\frac{\partial \widehat{\Omega}_{\mathrm{Sh}}^{\mathrm{ti}}}{\partial \lambda_a}
-
p\lambda_a^{-1},
\qquad
B_a
=
-
\frac{\partial \widehat{\Omega}_{\mathrm{Sh}}^{\mathrm{ti}}}{\partial H_a},
\qquad
a=1,2,3.
\end{equation}
Here, \(P_a\) are the principal nominal stress components, \(B_a\) are the material magnetic induction components in the principal basis. The advantage of this formulation is that it gives a compact and physically transparent representation of transversely isotropic magnetoelasticity, while allowing classical invariant-based models to be re-expressed in spectral form.

\subsubsection{\textbf{Anisotropic viscoelastic constitutive models}}\label{constvmodelanisoviscoelas}

\paragraph{\textbf{Saxena--Hossain--Steinmann transversely isotropic magneto-viscoelastic model}}
Saxena, Hossain and Steinmann extended the finite-strain isotropic magneto-viscoelastic framework to transversely isotropic magneto-active polymers with an aligned particle-chain microstructure \cite{saxena3}. The anisotropy is introduced by particle chains formed during curing under an applied magnetic field. In the present notation, the unit chain direction in the reference configuration is denoted by \(\boldsymbol A\), and the associated structure tensor is
\begin{equation}
\boldsymbol G
=
\boldsymbol A\otimes\boldsymbol A .
\end{equation}

The model separates the response into matrix and chain contributions. The matrix part retains the equilibrium and non-equilibrium structure of the isotropic magneto-viscoelastic theory, while the chain part introduces additional stretch and magnetic-induction variables along the preferred direction. The total stored-energy density is written as
\begin{equation}
\Omega
=
\Omega_{\infty}(\boldsymbol C,\mathbbm B)
+
\Omega_{\infty}^{c}(\boldsymbol C^{c},B^{c})
+
\Omega_{\mathrm v}(\boldsymbol C,\boldsymbol C_{\mathrm v},\mathbbm B,\mathbbm B_{\mathrm v})
+
\Omega_{\mathrm v}^{c}(\boldsymbol C^{c},\boldsymbol C_{\mathrm v}^{c},B^{c},B_{\mathrm v}^{c}).
\end{equation}
Here, \(\Omega_{\infty}\) and \(\Omega_{\mathrm v}\) are the equilibrium and non-equilibrium matrix contributions, while \(\Omega_{\infty}^{c}\) and \(\Omega_{\mathrm v}^{c}\) are the corresponding chain-level contributions. The superscript \(c\) denotes quantities associated with the chain direction, and the subscript \(\mathrm v\) denotes non-equilibrium internal variables.

The matrix kinematics are described by
\begin{equation}
\boldsymbol C
=
\boldsymbol F^{T}\boldsymbol F,
\qquad
\boldsymbol F
=
\boldsymbol F_{\mathrm e}\boldsymbol F_{\mathrm v},
\qquad
\boldsymbol C_{\mathrm v}
=
\boldsymbol F_{\mathrm v}^{T}\boldsymbol F_{\mathrm v}.
\end{equation}
The magnetic induction is decomposed additively as
\begin{equation}
\mathbbm B
=
\mathbbm B_{\mathrm e}
+
\mathbbm B_{\mathrm v},
\qquad
\mathbbm B_{\mathrm e}
=
\mathbbm B-\mathbbm B_{\mathrm v}.
\end{equation}
Thus, \(\boldsymbol C_{\mathrm v}\) governs mechanical relaxation of the matrix, while \(\mathbbm B_{\mathrm v}\) governs magnetic relaxation.

The chain deformation gradient is obtained by projecting the deformation along the particle-chain direction:
\begin{equation}
\boldsymbol F^{c}
=
\boldsymbol F\boldsymbol G .
\end{equation}
The corresponding chain right Cauchy--Green tensor is
\begin{equation}
\boldsymbol C^{c}
=
(\boldsymbol F^{c})^{T}\boldsymbol F^{c}
=
\boldsymbol G\boldsymbol C\boldsymbol G .
\end{equation}
The chain stretch and viscous chain stretch are introduced through
\begin{equation}
\boldsymbol C^{c}
=
(\lambda^{c})^{2}\boldsymbol G,
\qquad
\boldsymbol C_{\mathrm v}^{c}
=
(\lambda_{\mathrm v}^{c})^{2}\boldsymbol G,
\qquad
\lambda_{\mathrm e}^{c}
=
\frac{\lambda^{c}}{\lambda_{\mathrm v}^{c}} .
\end{equation}

The magnetic induction component along the chain direction is written as
\begin{equation}
\mathbbm B^{c}
=
B^{c}\boldsymbol A .
\end{equation}
Its equilibrium and non-equilibrium parts are
\begin{equation}
B^{c}
=
B_{\mathrm e}^{c}
+
B_{\mathrm v}^{c},
\qquad
B_{\mathrm e}^{c}
=
B^{c}-B_{\mathrm v}^{c}.
\end{equation}
Here, \(B^{c}\) is the scalar magnetic-induction component along \(\boldsymbol A\), while \(B_{\mathrm v}^{c}\) is its non-equilibrium counterpart.

The equilibrium matrix energy is taken in a Mooney--Rivlin-type magnetoelastic form:
\begin{equation}
\Omega_{\infty}
=
\frac{\mu_{\infty}}{4}
\left[
(1+n)(I_1-3)
+
(1-n)(I_2-3)
\right]
+
\frac{a_{\infty}}{2}I_4
+
\frac{b_{\infty}}{2}I_6 .
\end{equation}
The matrix invariants are
\begin{equation}
I_4
=
\mathbbm B\cdot\mathbbm B,
\qquad
I_6
=
(\boldsymbol C\mathbbm B)\cdot(\boldsymbol C\mathbbm B).
\end{equation}
The parameters \(\mu_{\infty}\), \(n\), \(a_{\infty}\), and \(b_{\infty}\) describe the equilibrium matrix shear response, Mooney--Rivlin weighting, magnetic storage, and magneto-mechanical coupling.

The non-equilibrium matrix energy is written as
\begin{equation}
\Omega_{\mathrm v}
=
\frac{\mu_{\mathrm v}}{2}
\left[
\boldsymbol C_{\mathrm v}^{-1}:\boldsymbol C
-
3
\right]
+
\frac{a_{\mathrm v}}{2}
\mathbbm B_{\mathrm e}\cdot\mathbbm B_{\mathrm e}
+
\frac{b_{\mathrm v}}{2}
(\boldsymbol C\mathbbm B_{\mathrm e})\cdot(\boldsymbol C\mathbbm B_{\mathrm e}) .
\end{equation}
Here, \(\mu_{\mathrm v}\), \(a_{\mathrm v}\), and \(b_{\mathrm v}\) are the corresponding non-equilibrium matrix parameters.

The equilibrium chain energy is taken as
\begin{equation}
\Omega_{\infty}^{c}
=
\frac{\mu_{\infty}^{c}}{2}
\left[
(\lambda^{c})^2
+
\frac{2}{\lambda^{c}}
-
3
\right]
+
\frac{b_{\infty}^{c}}{2}
(\lambda^{c})^4
(B^{c})^2 .
\end{equation}
The first term describes stretch along the chain direction, while the second term couples the chain stretch with the magnetic induction component along the same direction.

The non-equilibrium chain energy is
\begin{equation}
\Omega_{\mathrm v}^{c}
=
\frac{\mu_{\mathrm v}^{c}}{2}
\left[
(\lambda_{\mathrm e}^{c})^2
+
\frac{2}{\lambda_{\mathrm e}^{c}}
-
3
\right]
+
\frac{b_{\mathrm v}^{c}}{2}
(\lambda^{c})^4
(B_{\mathrm e}^{c})^2 .
\end{equation}
Here, \(\mu_{\mathrm v}^{c}\) and \(b_{\mathrm v}^{c}\) are the chain-level non-equilibrium mechanical and magnetoelastic parameters.

Thermodynamic admissibility requires
\begin{equation}
\frac{\partial\Omega}{\partial\boldsymbol C_{\mathrm v}}
:
\dot{\boldsymbol C}_{\mathrm v}
+
\frac{\partial\Omega}{\partial\mathbbm B_{\mathrm v}}
\cdot
\dot{\mathbbm B}_{\mathrm v}
+
\frac{\partial\Omega}{\partial\lambda_{\mathrm v}^{c}}
\dot{\lambda}_{\mathrm v}^{c}
+
\frac{\partial\Omega}{\partial B_{\mathrm v}^{c}}
\dot B_{\mathrm v}^{c}
\leq 0 .
\end{equation}
For the matrix variables, the evolution equations are chosen as
\begin{equation}
\dot{\boldsymbol C}_{\mathrm v}
=
\frac{1}{T_{\mathrm v}}
\left[
\boldsymbol C
-
\frac{1}{3}
\left(
\boldsymbol C:\boldsymbol C_{\mathrm v}^{-1}
\right)
\boldsymbol C_{\mathrm v}
\right],
\end{equation}
and
\begin{equation}
\dot{\mathbbm B}_{\mathrm v}
=
\frac{\mu_0}{T_{\mathrm m}}
\left[
a_{\mathrm v}\boldsymbol I
+
b_{\mathrm v}\boldsymbol C^2
\right]
(\mathbbm B-\mathbbm B_{\mathrm v}) .
\end{equation}
Here, \(T_{\mathrm v}\) and \(T_{\mathrm m}\) are the matrix mechanical and magnetic relaxation times.

For the chain magnetic internal variable, the relaxation law is written with the normalization consistent with the energy above:
\begin{equation}
\dot B_{\mathrm v}^{c}
=
-\frac{1}{b_{\mathrm v}^{c}T_{\mathrm m}^{c}}
\frac{\partial\Omega_{\mathrm v}^{c}}{\partial B_{\mathrm v}^{c}}
=
\frac{1}{T_{\mathrm m}^{c}}
(\lambda^{c})^{4}
\left(
B^{c}-B_{\mathrm v}^{c}
\right).
\end{equation}
Similarly, the chain-stretch internal variable evolves according to
\begin{equation}
\dot{\lambda}_{\mathrm v}^{c}
=
-\frac{\lambda_{\mathrm v}^{c}}{\mu_{\mathrm v}^{c}T_{\mathrm v}^{c}}
\frac{\partial\Omega_{\mathrm v}^{c}}{\partial\lambda_{\mathrm v}^{c}}
=
\frac{1}{T_{\mathrm v}^{c}}
\left[
\frac{(\lambda^{c})^2}{(\lambda_{\mathrm v}^{c})^2}
-
\frac{\lambda_{\mathrm v}^{c}}{\lambda^{c}}
\right].
\end{equation}
The parameters \(T_{\mathrm m}^{c}\) and \(T_{\mathrm v}^{c}\) are the magnetic and mechanical relaxation times associated with the chain response. At equilibrium,
\begin{equation}
\boldsymbol C_{\mathrm v}
=
\boldsymbol C,
\qquad
\mathbbm B_{\mathrm v}
=
\mathbbm B,
\qquad
\lambda_{\mathrm v}^{c}
=
\lambda^{c},
\qquad
B_{\mathrm v}^{c}
=
B^{c}.
\end{equation}

The total second Piola--Kirchhoff stress is then obtained from the energy as
\begin{equation}
\boldsymbol S^{\mathrm{tot}}
=
-p\boldsymbol C^{-1}
+
2\frac{\partial\Omega_{\infty}}{\partial\boldsymbol C}
+
2\frac{\partial\Omega_{\mathrm v}}{\partial\boldsymbol C}
+
2\boldsymbol G
\frac{\partial\Omega_{\infty}^{c}}{\partial\boldsymbol C^{c}}
\boldsymbol G
+
2\boldsymbol G
\frac{\partial\Omega_{\mathrm v}^{c}}{\partial\boldsymbol C^{c}}
\boldsymbol G .
\end{equation}
The material magnetic field follows from differentiation with respect to the magnetic induction:
\begin{equation}
\mathbbm H
=
\frac{\partial\Omega_{\infty}}{\partial\mathbbm B}
+
\frac{\partial\Omega_{\mathrm v}}{\partial\mathbbm B}
+
\frac{1}{(\lambda^{c})^2}
\left[
\frac{\partial\Omega_{\infty}^{c}}{\partial B^{c}}
+
\frac{\partial\Omega_{\mathrm v}^{c}}{\partial B^{c}}
\right]
\boldsymbol C\boldsymbol A .
\end{equation}
The first two terms are the matrix contributions, while the final term is the chain contribution mapped back to the material configuration.
Thus, the model provides a transversely isotropic extension of finite-strain magneto-viscoelasticity in which matrix relaxation and chain-direction relaxation are treated separately but consistently within the same thermodynamic framework.

\subsection{Constitutive models for hard-magnetic MAPs}\label{hardmagsoftmaterials}

Having introduced the distinction between soft-magnetic and hard-magnetic soft materials in the Introduction, this subsection focuses on constitutive models in which a prescribed or evolving remanent magnetic state plays the central role. In these materials, the retained magnetic state interacts with the applied magnetic field and gives rise to magnetic torques, bending, twisting and programmed shape transformations. The models reviewed below are arranged chronologically. They differ mainly in how the remanent magnetic quantity is represented, whether the response is treated as purely energetic or dissipative, and whether the magnetic variables are formulated in an \(\boldsymbol F-\mathbbm H\) or \(\boldsymbol F-\mathbbm B\) setting.




\subsubsection{\textbf{Zhao--Kim--Chester--Sharma--Zhao ideal hard-magnetic soft-material model}}

Zhao et al. \cite{zhao2019mechanics} proposed one of the first widely used continuum models for hard-magnetic soft materials with prescribed residual magnetic flux density. The model is particularly suited to structures with programmed magnetic domains, such as beams, plates and printed shape-morphing systems. Its central assumption is that, after magnetic saturation, the material retains a residual magnetic flux density, and the applied actuation field remains well below the coercive field strength. In this regime, the magnetic response is idealised by
\begin{equation}
\mathbbm h
=
\frac{1}{\mu_0}
\left(
\mathbbm b-\mathbbm b_r
\right),
\label{eq:zhao_h_b_relation}
\end{equation}
where \(\mathbbm b_r\) is the spatial residual magnetic flux density and \(\mu_0\) is the vacuum permeability. The corresponding material residual flux density is defined through
\begin{equation}
\mathbbm B_r
=
J\boldsymbol F^{-1}\mathbbm b_r,
\qquad
\mathbbm b_r
=
J^{-1}\boldsymbol F\mathbbm B_r .
\label{eq:zhao_residual_flux_relation}
\end{equation}

For a prescribed applied magnetic flux density \(\mathbbm b^{\mathrm{app}}\), the nominal Helmholtz free-energy density is written as
\begin{equation}
\Omega(\boldsymbol F)
=
\Omega^{\mathrm e}(\boldsymbol F)
+
\Omega^{\mathrm{mag}}(\boldsymbol F,\mathbbm B_r,\mathbbm b^{\mathrm{app}}),
\label{eq:zhao_total_energy}
\end{equation}
with
\begin{equation}
\Omega^{\mathrm{mag}}
=
-
\frac{1}{\mu_0}
\left(
\boldsymbol F\mathbbm B_r
\right)\cdot\mathbbm b^{\mathrm{app}} .
\label{eq:zhao_magnetic_energy}
\end{equation}
Thus,
\begin{equation}
\Omega(\boldsymbol F)
=
\Omega^{\mathrm e}(\boldsymbol F)
-
\frac{1}{\mu_0}
\left(
\boldsymbol F\mathbbm B_r
\right)\cdot\mathbbm b^{\mathrm{app}} .
\label{eq:zhao_energy}
\end{equation}
A commonly used compressible neo-Hookean elastic part is
\begin{equation}
\Omega^{\mathrm e}(\boldsymbol F)
=
\frac{\mu}{2}
\left(
J^{-2/3}I_1-3
\right)
+
\frac{\kappa}{2}
\left(
J-1
\right)^2.
\label{eq:zhao_compressible_neohookean}
\end{equation}

The total first Piola stress follows as
\begin{equation}
\boldsymbol P^{\mathrm{tot}}
=
\frac{\partial\Omega^{\mathrm e}}{\partial\boldsymbol F}
-
\frac{1}{\mu_0}
\mathbbm b^{\mathrm{app}}\otimes\mathbbm B_r .
\label{eq:zhao_piola}
\end{equation}
The corresponding Cauchy stress is
\begin{equation}
\boldsymbol\sigma^{\mathrm{tot}}
=
\frac{1}{J}
\frac{\partial\Omega^{\mathrm e}}{\partial\boldsymbol F}
\boldsymbol F^T
-
\frac{1}{\mu_0}
\mathbbm b^{\mathrm{app}}\otimes\mathbbm b_r .
\label{eq:zhao_cauchy}
\end{equation}
This model is simple, efficient and effective for many programmed slender structures. Its magnetic stress contribution may be non-symmetric when the applied field is not aligned with the residual magnetic flux density, and this non-symmetric part is associated with the magnetic torque that drives bending, twisting and shape transformation.

\subsubsection{\textbf{Mukherjee--Rambausek--Danas dissipative model}}

Mukherjee, Rambausek and Danas \cite{mukherjee2021explicit} developed a thermodynamically consistent dissipative model for isotropic incompressible hard-magnetic particle-filled elastomers. Unlike the ideal model of Zhao et al. \cite{zhao2019mechanics}, this formulation treats the remanent magnetic state as an internal variable and is designed to account for magnetic hysteresis in the hard particles. The internal variable is denoted by \(\boldsymbol\xi\), and it represents the evolving remanent magnetic state in an intermediate configuration.

The magnetic invariants involving the material magnetic field \(\mathbbm H\), the tensor \(\boldsymbol C\), and the internal variable \(\boldsymbol\xi\) are
\begin{equation}
I_5
=
\mathbbm H\cdot\boldsymbol C^{-1}\mathbbm H,
\qquad
I_5^{er}
=
\mathbbm H\cdot\boldsymbol C^{-1/2}\boldsymbol\xi,
\qquad
I_5^r
=
\boldsymbol\xi\cdot\boldsymbol\xi ,
\label{eq:mrd_I5_invariants}
\end{equation}
and
\begin{equation}
I_4
=
\mathbbm H\cdot\mathbbm H,
\qquad
I_4^{er}
=
\mathbbm H\cdot\boldsymbol C^{1/2}\boldsymbol\xi,
\qquad
I_4^r
=
\boldsymbol\xi\cdot\boldsymbol C\boldsymbol\xi .
\label{eq:mrd_I4_invariants}
\end{equation}
Here, \(I_5\) is the magnetoelastic invariant associated with the pull-back of the spatial magnetic field, \(I_5^{er}\) and \(I_4^{er}\) couple the magnetic field with the remanent internal variable, while \(I_5^r\) and \(I_4^r\) measure the internal magnetic state before and after the action of \(\boldsymbol C\).

For an incompressible material, the Helmholtz free-energy density is written as
\begin{equation}
\Omega
=
\begin{cases}
\rho_0\Psi^{\mathrm{mech}}(I_1)
+
\rho_0\Psi^{\mathrm{mag}}(I_5,I_5^{er},I_5^r)
+
\rho_0\Psi^{\mathrm{couple}}(I_4^{er},I_4^r,I_5^{er},I_5^r)
-
\dfrac{\mu_0}{2}I_5,
& J=1,\\[3mm]
+\infty,
& J\neq 1 .
\end{cases}
\label{eq:mrd_total_energy}
\end{equation}
The mechanical part is taken as
\begin{equation}
\rho_0\Psi^{\mathrm{mech}}(I_1)
=
(1-c)\rho_0\Psi_m^{\mathrm{mech}}(\mathcal I_1),
\qquad
\mathcal I_1
=
\frac{I_1-3}{(1-c)^{7/2}}+3,
\label{eq:mrd_mechanical_energy}
\end{equation}
where \(c\) is the particle volume fraction and \(\Psi_m^{\mathrm{mech}}\) is the matrix free-energy density. For a neo-Hookean matrix,
\begin{equation}
\rho_0\Psi_m^{\mathrm{mech}}(\mathcal I_1)
=
\frac{G_m}{2}
\left(
\mathcal I_1-3
\right),
\label{eq:mrd_matrix_neohookean}
\end{equation}
where \(G_m\) is the matrix shear modulus.

The magnetic contribution is decomposed into energetic and remanent parts,
\begin{equation}
\Psi^{\mathrm{mag}}
=
\Psi^{e,\mathrm{mag}}
+
\Psi^{r,\mathrm{mag}}.
\label{eq:mrd_mag_decomp}
\end{equation}
The energetic magnetic contribution is
\begin{equation}
\rho_0\Psi^{e,\mathrm{mag}}(I_5)
=
-
\frac{\mu_0}{2}\chi^e I_5,
\qquad
\chi^e
=
\frac{3c\chi_p^e}{3+(1-c)\chi_p^e},
\label{eq:mrd_energetic_magnetic}
\end{equation}
where \(\chi_p^e\) is the energetic susceptibility of the particles and \(\chi^e\) is the corresponding effective susceptibility. The remanent magnetic contribution is
\begin{equation}
\rho_0\Psi^{r,\mathrm{mag}}
=
\mu_0(1+\chi^e)I_5^{er}
+
\frac{\mu_0}{2}
\left[
\frac{1-c}{3c\chi_p^r}I_5^r
+
\frac{(m^s)^2}{c\chi_p^r}
f_p
\left(
\frac{\sqrt{I_5^r}}{m^s}
\right)
\right],
\label{eq:mrd_remanent_magnetic}
\end{equation}
where \(\chi_p^r\) is the remanent susceptibility of the particles and \(m^s\) is the effective saturation magnetisation,
\begin{equation}
m^s
=
c\,m_p^s
\left(
\frac{1+\chi_p^e}{1+\chi^e}
\right)^{q_{\mathrm{sat}}}.
\label{eq:mrd_saturation}
\end{equation}
A saturation-type response may be represented by
\begin{equation}
f_p(x)
=
-
\left[
\log(1-x)+x
\right].
\label{eq:mrd_saturation_function}
\end{equation}

The magneto-mechanical coupling energy is
\begin{equation}
\rho_0\Psi^{\mathrm{couple}}
=
c\beta(c)\mu_0
\left[
\left(
I_4^r-I_5^r
\right)
-
2\chi^e
\left(
I_4^{er}-I_5^{er}
\right)
\right],
\label{eq:mrd_coupling_energy}
\end{equation}
where a fitted coupling function is
\begin{equation}
\beta(c)
=
19c^2-10.4c+1.71 .
\label{eq:mrd_beta}
\end{equation}

The evolution of the internal variable follows the generalized standard material structure,
\begin{equation}
\frac{\partial\Omega}{\partial\boldsymbol\xi}
+
\frac{\partial D}{\partial\dot{\boldsymbol\xi}}
=
\boldsymbol 0 ,
\label{eq:mrd_evolution}
\end{equation}
where \(D\) is the dissipation potential. A rate-dependent choice is
\begin{equation}
D(\dot{\boldsymbol\xi})
=
\frac{n}{n+1}
b_c
\left|
\dot{\boldsymbol\xi}
\right|^{(n+1)/n},
\qquad
1\leq n<\infty ,
\label{eq:mrd_rate_dependent_dissipation}
\end{equation}
with \(b_c\) denoting the effective coercive field. The rate-independent limit is
\begin{equation}
D(\dot{\boldsymbol\xi})
=
b_c
\left|
\dot{\boldsymbol\xi}
\right|,
\label{eq:mrd_rate_independent_dissipation}
\end{equation}
where
\begin{equation}
b_c
=
b_p^c
\left(
\frac{1+\chi^e}{1+\chi_p^e}
\right)^{q_{\mathrm{coer}}},
\label{eq:mrd_coercive_field}
\end{equation}
and \(b_p^c\) is the particle coercive field. The switching surface is
\begin{equation}
\Phi(\mathbbm E^r)
=
\left|
\mathbbm E^r
\right|^2-b_c^2
=
0,
\label{eq:mrd_switching_surface}
\end{equation}
where \(\mathbbm E^r\) is the thermodynamic force conjugate to \(\boldsymbol\xi\). The associated flow rule and Kuhn--Tucker conditions are
\begin{equation}
\dot{\boldsymbol\xi}
=
\dot\Lambda
\frac{\partial\Phi}{\partial\mathbbm E^r},
\qquad
\dot\Lambda\geq 0,
\qquad
\Phi\leq 0,
\qquad
\dot\Lambda\Phi=0 .
\label{eq:mrd_kuhn_tucker}
\end{equation}

After the internal variable has been updated, the total second Piola stress and material magnetic induction are obtained from
equations \eqref{const_rel_mag} and \eqref{incomp_const}.
This formulation is therefore suited to cases where magnetic history and hysteretic switching of the hard particles are important.

\subsubsection{\textbf{Mukherjee--Danas unified dual formulation}}

Mukherjee and Danas \cite{mukherjee2022unified} extended the preceding dissipative framework into a unified dual formulation that can be written either in an \(\boldsymbol F-\mathbbm H\) setting or in an \(\boldsymbol F-\mathbbm B\) setting. The two descriptions are related by a partial Legendre--Fenchel transformation with respect to the magnetic variables. This is useful because scalar-potential and vector-potential finite-element formulations naturally use different magnetic primary variables.

In the \(\boldsymbol F-\mathbbm H\) setting, the energy is
\begin{equation}
\Omega
=
\Omega^{\mathbbm H}
(\boldsymbol C,\mathbbm H,\boldsymbol\xi),
\label{eq:dual_h_energy_general}
\end{equation}
with
\begin{equation}
\Omega^{\mathbbm H}
=
\begin{cases}
\rho_0\Psi^{\mathrm{mech}}(I_1)
+
\rho_0\Psi_{\mathbbm H}^{\mathrm{mag}}
+
\rho_0\Psi_{\mathbbm H}^{\mathrm{couple}}
-
\dfrac{\mu_0}{2}I_5^{\mathbbm H\mathbbm H},
& J=1,\\[3mm]
+\infty,
& J\neq 1 .
\end{cases}
\label{eq:dual_h_energy}
\end{equation}
The invariants are
\begin{equation}
I_5^{\mathbbm H\mathbbm H}
=
\mathbbm H\cdot\boldsymbol C^{-1}\mathbbm H,
\qquad
I_5^{\mathbbm H\xi}
=
\mathbbm H\cdot\boldsymbol C^{-1/2}\boldsymbol\xi,
\qquad
I_5^{\xi\xi}
=
\boldsymbol\xi\cdot\boldsymbol\xi,
\label{eq:dual_h_i5}
\end{equation}
and
\begin{equation}
I_4^{\mathbbm H\xi}
=
\mathbbm H\cdot\boldsymbol C^{1/2}\boldsymbol\xi,
\qquad
I_4^{\xi\xi}
=
\boldsymbol\xi\cdot\boldsymbol C\boldsymbol\xi .
\label{eq:dual_h_i4}
\end{equation}
Here, \(I_5^{\mathbbm H\mathbbm H}\) is the magnetic-field invariant in the \(\boldsymbol F-\mathbbm H\) setting, \(I_5^{\mathbbm H\xi}\) and \(I_4^{\mathbbm H\xi}\) couple the magnetic field to the remanent internal variable, and \(I_5^{\xi\xi}\) and \(I_4^{\xi\xi}\) describe the internal magnetic state.

The magnetic contribution is
\begin{equation}
\rho_0\Psi_{\mathbbm H}^{\mathrm{mag}}
=
-
\frac{\mu_0}{2}\chi^e I_5^{\mathbbm H\mathbbm H}
+
\mu_0(1+\chi^e)I_5^{\mathbbm H\xi}
+
\frac{\mu_0}{2}
\left[
\frac{1-c}{3c\chi_p^r}I_5^{\xi\xi}
+
\frac{(m^s)^2}{c\chi_p^r}
f_p
\left(
\frac{\sqrt{I_5^{\xi\xi}}}{m^s}
\right)
\right],
\label{eq:dual_h_mag_energy}
\end{equation}
and the coupling contribution is
\begin{equation}
\rho_0\Psi_{\mathbbm H}^{\mathrm{couple}}
=
c\beta(c)\mu_0
\left[
\left(
I_4^{\xi\xi}-I_5^{\xi\xi}
\right)
-
2\chi^e
\left(
I_4^{\mathbbm H\xi}-I_5^{\mathbbm H\xi}
\right)
\right].
\label{eq:dual_h_coupling_energy}
\end{equation}

The evolution equations for \(\boldsymbol\xi\) are the same as those in the dissipative hard-magnetic model. Once the internal variable has been updated, the constitutive relations in the \(\boldsymbol F-\mathbbm H\) setting are
\begin{equation}
\boldsymbol S^{\mathrm{tot}}
=
2\frac{\partial\Omega^{\mathbbm H}}{\partial\boldsymbol C}
-
p\boldsymbol C^{-1},
\qquad
\mathbbm B
=
-
\frac{\partial\Omega^{\mathbbm H}}{\partial\mathbbm H}.
\label{eq:dual_h_constitutive}
\end{equation}

The equivalent \(\boldsymbol F-\mathbbm B\) energy is obtained through
\begin{equation}
\Omega^{\mathbbm B}
(\boldsymbol C,\mathbbm B,\boldsymbol\xi)
=
\sup_{\mathbbm H}
\left[
\Omega^{\mathbbm H}
(\boldsymbol C,\mathbbm H,\boldsymbol\xi)
+
\mathbbm H\cdot\mathbbm B
\right].
\label{eq:dual_legendre}
\end{equation}
In this setting,
\begin{equation}
\Omega^{\mathbbm B}
=
\rho_0\Psi^{\mathrm{mech}}
+
\rho_0\Psi_{\mathbbm B}^{\mathrm{mag}}
+
\rho_0\Psi_{\mathbbm B}^{\mathrm{couple}}
+
\frac{1}{2\mu_0}I_5^{\mathbbm B}.
\label{eq:dual_b_energy}
\end{equation}
The corresponding invariants are
\begin{equation}
I_5^{\mathbbm B}
=
\mathbbm B\cdot\boldsymbol C\mathbbm B,
\qquad
I_5^{\mathbbm B\xi}
=
\mathbbm B\cdot\boldsymbol C^{1/2}\boldsymbol\xi,
\qquad
I_5^{\xi\xi}
=
\boldsymbol\xi\cdot\boldsymbol\xi,
\label{eq:dual_b_invariants_1}
\end{equation}
and
\begin{equation}
I_4^{\xi\xi}
=
\boldsymbol\xi\cdot\boldsymbol C\boldsymbol\xi,
\qquad
I_6^{\mathbbm B\xi}
=
\mathbbm B\cdot\boldsymbol C^{3/2}\boldsymbol\xi .
\label{eq:dual_b_invariants_2}
\end{equation}
Here, \(I_5^{\mathbbm B}\) is the magnetic-induction invariant, \(I_5^{\mathbbm B\xi}\) and \(I_6^{\mathbbm B\xi}\) couple the magnetic induction with the remanent internal variable, and \(I_4^{\xi\xi}\) and \(I_5^{\xi\xi}\) describe the internal magnetic state.

The coupling energy is
\begin{equation}
\rho_0\Psi_{\mathbbm B}^{\mathrm{couple}}
=
c\beta(c)
\left[
\mu_0(1-2\chi^e)
\left(
I_4^{\xi\xi}-I_5^{\xi\xi}
\right)
-
\frac{2\chi^e}{1+\chi^e}
\left(
I_6^{\mathbbm B\xi}-I_5^{\mathbbm B\xi}
\right)
\right].
\label{eq:dual_b_coupling_energy}
\end{equation}

After the internal variable has been updated, the constitutive relations in the \(\boldsymbol F-\mathbbm B\) setting are
\begin{equation}
\boldsymbol S^{\mathrm{tot}}
=
2\frac{\partial\Omega^{\mathbbm B}}{\partial\boldsymbol C}
-
p\boldsymbol C^{-1},
\qquad
\mathbbm H
=
\frac{\partial\Omega^{\mathbbm B}}{\partial\mathbbm B}.
\label{eq:dual_b_constitutive}
\end{equation}
This dual formulation provides equivalent descriptions in the two magnetic variable settings while retaining the same internal variable and dissipative structure.

\subsubsection{\textbf{Danas--Reis stretch-independent magnetisation model}}

Danas and Reis \cite{danas2024stretch} revisited the modelling of incompressible isotropic hard-magnetic particle-filled elastomers after full pre-magnetisation and clarified the role of stretch in the magnetisation response. Their analysis shows that, after full pre-magnetisation, the experimentally relevant magnetisation response can be stretch-independent. This behaviour is captured when the current magnetisation is related to the rotational part of the deformation rather than to the full deformation gradient.

In the \(\boldsymbol F-\mathbbm B\) setting, the energy density may be written as
\begin{equation}
\Omega(\boldsymbol F,\mathbbm B,\mathbbm H^r)
=
\rho_0\Psi^{\mathrm{mech}}(I_1,J)
+
\rho_0\Psi^{\mathrm{mag}}
\left(
I_5^{\mathbbm B},
I_5^{\mathbbm B H^r},
I_5^{H^r}
\right)
+
\rho_0\Psi^{\mathrm{couple}}
\left(
I_4^{H^r},
I_5^{\mathbbm B H^r},
I_5^{H^r},
I_6^{\mathbbm B H^r}
\right)
+
\frac{1}{2\mu_0}I_5^{\mathbbm B}.
\label{eq:danas_reis_total_energy}
\end{equation}
The invariants are
\begin{equation}
I_5^{\mathbbm B}
=
\mathbbm B\cdot\boldsymbol C\mathbbm B,
\qquad
I_5^{\mathbbm B H^r}
=
\mathbbm B\cdot\boldsymbol C^{1/2}\mathbbm H^r,
\qquad
I_5^{H^r}
=
\mathbbm H^r\cdot\mathbbm H^r,
\label{eq:danas_reis_i5}
\end{equation}
and
\begin{equation}
I_4^{H^r}
=
\mathbbm H^r\cdot\boldsymbol C\mathbbm H^r,
\qquad
I_6^{\mathbbm B H^r}
=
\mathbbm B\cdot\boldsymbol C^{3/2}\mathbbm H^r .
\label{eq:danas_reis_i4_i6}
\end{equation}
Here, \(\mathbbm H^r\) is the remanent magnetic internal variable used in this formulation. The invariant \(I_5^{\mathbbm B}\) measures the magnetic induction through \(\boldsymbol C\), \(I_5^{\mathbbm B H^r}\) and \(I_6^{\mathbbm B H^r}\) couple \(\mathbbm B\) with \(\mathbbm H^r\), while \(I_5^{H^r}\) and \(I_4^{H^r}\) describe the remanent state before and after the action of \(\boldsymbol C\).

The mechanical energy can be written as
\begin{equation}
\rho_0\Psi^{\mathrm{mech}}(I_1,J)
=
(1-c)\rho_0\Psi_m^{\mathrm{mech}}(\mathcal I_1)
+
\frac{K_m}{2(1-c)^6}
\left(
J-1
\right)^2,
\qquad
\mathcal I_1
=
\frac{J^{-2/3}I_1-3}{(1-c)^{7/2}}+3 .
\label{eq:danas_reis_mechanical_energy}
\end{equation}
The magnetic energy is
\begin{equation}
\rho_0\Psi^{\mathrm{mag}}
=
-
\frac{1}{2\mu_0}
\frac{\chi^e}{1+\chi^e}
I_5^{\mathbbm B}
+
I_5^{\mathbbm B H^r}
+
\frac{\mu_0}{2}
\left(
\chi^e+\frac{1+2c}{3c\chi_p^r}
\right)
I_5^{H^r}
+
\frac{\mu_0(m^s)^2}{c\chi_p^r}
f_p
\left(
\frac{\sqrt{I_5^{H^r}}}{m^s}
\right),
\label{eq:danas_reis_magnetic_energy}
\end{equation}
with
\begin{equation}
\chi^e
=
\frac{3c\chi_p^e}{3+(1-c)\chi_p^e},
\qquad
m^s
=
c\,m_p^s
\left(
\frac{1+\chi_p^e}{1+\chi^e}
\right).
\label{eq:danas_reis_effective_parameters}
\end{equation}
A suitable saturation function is
\begin{equation}
f_p(x)
=
-
\left[
\log(1-x)+x
\right].
\label{eq:danas_reis_saturation_function}
\end{equation}
The coupling energy is
\begin{equation}
\rho_0\Psi^{\mathrm{couple}}
=
c\beta(c)
\left[
\mu_0(1-2\chi^e)
\left(
I_4^{H^r}-I_5^{H^r}
\right)
-
\frac{2\chi^e}{1+\chi^e}
\left(
I_6^{\mathbbm B H^r}-I_5^{\mathbbm B H^r}
\right)
\right],
\label{eq:danas_reis_coupling_energy}
\end{equation}
where
\begin{equation}
\beta(c)
=
19c^2-10.4c+1.71 .
\label{eq:danas_reis_beta}
\end{equation}

The rate-independent dissipation potential is
\begin{equation}
D(\dot{\mathbbm H}^r)
=
b_c
\left|
\dot{\mathbbm H}^r
\right|,
\qquad
b_c
=
b_p^c
\left(
\frac{1+\chi^e}{1+\chi_p^e}
\right)^{4/5}.
\label{eq:danas_reis_dissipation}
\end{equation}
The evolution of \(\mathbbm H^r\) follows
\begin{equation}
\frac{\partial\Omega}{\partial\mathbbm H^r}
+
\frac{\partial D}{\partial\dot{\mathbbm H}^r}
=
\boldsymbol 0 .
\label{eq:danas_reis_evolution}
\end{equation}

After the remanent internal variable has been determined, the constitutive relations are same as the {Mukherjee--Danas unified dual formulation model in the \(\boldsymbol F-\mathbbm B\) setting.
For an already pre-magnetised body subjected to small actuation fields, the remanent magnetic state may be treated as fixed and the dissipative evolution can be suppressed. If \(\boldsymbol F=\boldsymbol R\boldsymbol U\) is the polar decomposition, a stretch-independent representation of the current magnetisation is written as
\begin{equation}
\mathbbm m
=
J^{-1}\boldsymbol R\mathbbm M .
\label{eq:danas_reis_rotation_magnetisation}
\end{equation}
This differs from a full \(\boldsymbol F\)-push-forward of the remanent magnetic quantity, which may introduce artificial stretch dependence. The distinction is especially important in problems involving significant stretching. For slender structures dominated by bending, where stretching is often small, the ideal model of Zhao et al. \cite{zhao2019mechanics} can still provide accurate structural predictions.

\subsubsection{\textbf{Lin--Hooshmand-Ahoor--Bodelot--Danas hard-magnetic foam model}}

Lin et al. \cite{lin2025foams} extended hard-magnetic soft-material modelling to mechanically soft, magnetically hard foams. This work is distinct from the dense incompressible formulations discussed above because the material is porous and highly compressible. The central idea is that deformation-induced pore closure, pore opening and specimen-shape change can alter the surrounding magnetic flux in a measurable way. This enables the material to be used not only for actuation, but also for deformation and haptic sensing.

The formulation is based on finite-strain compressible magnetoelasticity and combines experimental characterisation, analytical modelling and finite-element simulations. In a compact review notation, the Helmholtz free-energy density may be written as \begin{equation} \Omega = \Omega^{\mathrm{foam}} (\boldsymbol C,J,\mathbbm B,\mathbbm M_r;\Pi,\phi), \label{eq:foam_general_energy} \end{equation} where \(\Pi\) denotes the foam porosity, \(\phi\) is the magnetic-particle volume fraction, and \(\mathbbm M_r\) represents the remanent magnetic state of the hard-magnetic particle-filled skeleton. A useful schematic decomposition is \begin{equation} \Omega = \Omega^{\mathrm{mech}}(I_1,J;\Pi) + \Omega^{\mathrm{mag}}(\mathbbm B,\mathbbm M_r;\phi,\Pi) + \Omega^{\mathrm{couple}}(\boldsymbol C,J,\mathbbm B,\mathbbm M_r;\phi,\Pi), \label{eq:foam_decomposed_energy} \end{equation} where \(\Omega^{\mathrm{mech}}\) represents the compressible mechanical response of the porous skeleton, \(\Omega^{\mathrm{mag}}\) describes the magnetic contribution associated with the remanent magnetic state, and \(\Omega^{\mathrm{couple}}\) accounts for magneto-mechanical and magneto-dilatational coupling effects.

For compressible foams, the volumetric response plays a central role. Thus, in addition to the standard invariant $I_1$,
$J=\det\boldsymbol F$ is an essential constitutive argument. The magnetic invariants may be chosen consistently with the \(\boldsymbol F-\mathbbm B\) setting, for example
\begin{equation}
I_5^{\mathbbm B}
=
\mathbbm B\cdot\boldsymbol C\mathbbm B,
\qquad
I_5^{\mathbbm B M_r}
=
\mathbbm B\cdot\boldsymbol C^{1/2}\mathbbm M_r,
\qquad
I_4^{M_r}
=
\mathbbm M_r\cdot\boldsymbol C\mathbbm M_r .
\label{eq:foam_magnetic_invariants}
\end{equation}
Here, \(I_5^{\mathbbm B}\) measures the magnetic induction through the deformation, \(I_5^{\mathbbm B M_r}\) couples the magnetic induction with the remanent state, and \(I_4^{M_r}\) measures the deformation-dependent remanent magnetic contribution.

The total second Piola stress and magnetic field follow from
\begin{equation*}
\boldsymbol S^{\mathrm{tot}}
=
2\frac{\partial\Omega}{\partial\boldsymbol C},
\qquad
\mathbbm H
=
\frac{\partial\Omega}{\partial\mathbbm B}.
\label{eq:foam_constitutive}
\end{equation*}
The importance of this model is that it moves hard-magnetic soft-material modelling beyond dense incompressible elastomers and highlights the role of compressibility, porosity and shape-induced changes in magnetic flux. It is therefore particularly relevant for sensing-oriented applications, where deformation is inferred from measurable magnetic-field changes rather than being used only to generate motion.

\section{Conclusion and outlook}\label{concl}

\noindent This review has presented a structured survey of constitutive modelling approaches for magneto-active polymers at finite strains. The discussion has shown that MAP modelling has developed from early semi-empirical descriptions of field-induced stiffening into a broad family of thermodynamically consistent nonlinear continuum models capable of describing large deformation, magnetic coupling, anisotropy, remanent magnetic states and rate-dependent response \cite{jolly1,ginder1,ginder4,dorfmann3,dorfmann4,dorfmann5,zhao2019mechanics}. It has also emphasised that MAPs should be understood through two related but distinct material classes: soft-magnetic MAPs, in which the magnetisation is mainly induced by the applied field, and hard-magnetic MAPs, in which a programmed remanent magnetic state can drive shape transformation under comparatively small actuation fields \cite{kim2018printing,zhao2019mechanics}.

\noindent For soft-magnetic MAPs, the main constitutive developments include invariant-based finite-strain formulations, variational approaches, spectral representations, polyconvex frameworks, micromechanically motivated models, anisotropic energy functions and dispersed-chain descriptions \cite{dorfmann3,dorfmann4,kankanala1,bustamante1,danas1,shariff5,saxena4,ethirajMiehe2016}. These models show that the macroscopic response is governed not only by the polymer matrix and the applied magnetic field, but also by particle morphology, particle interactions, chain formation and the degree of anisotropy introduced during curing or loading. For hard-magnetic MAPs, recent models have introduced a complementary viewpoint in which the remanent magnetic state becomes a key constitutive ingredient. The ideal hard-magnetic MAP model of Zhao et al. \cite{zhao2019mechanics}, together with subsequent dissipative, dual and stretch-independent magnetisation models, provides a basis for describing programmed magnetic actuation, magnetic torque, magnetic hysteresis and large shape changes in soft structures.

\noindent The survey also shows that no single constitutive model is universally suitable for all MAPs and all loading conditions. The appropriate model depends on the material architecture, particle coercivity, deformation range, field intensity, degree of anisotropy, magnetic history, importance of rate effects and intended application. Simple semi-empirical models remain useful for preliminary interpretation and design. Invariant-based and variational theories provide rigorous finite-strain foundations. Spectral models improve physical transparency. Polyconvex formulations improve mathematical admissibility. Microstructural and dispersed-chain models provide closer links to particle morphology. Viscoelastic and internal-variable formulations are indispensable when dissipation, cyclic response, magnetic switching or transient actuation is experimentally significant.

\noindent Several challenges remain open. First, further experimentally validated constitutive models are needed for strongly coupled loading paths, large magnetic fields, saturation effects, remanent magnetic states, non-proportional deformation and combined mechanical and magnetic cycling. Second, the connection between macroscopic constitutive parameters and evolving particle-scale mechanisms requires deeper development, particularly for chain formation, chain breakage, particle interaction, filler-matrix debonding and microstructural rearrangement under repeated magnetic loading \cite{jolly1,hossain1,hossain2,saxena4,garciaGonzalezHossain2021}. Third, the modelling of hard-magnetic MAPs requires further refinement in situations involving large stretch, evolving remanence, viscoelasticity, local demagnetisation and complex three-dimensional magnetic programming. Fourth, thermo-magneto-mechanical coupling remains comparatively underdeveloped and should be integrated more systematically with finite-strain magnetoelastic, remanent-field and magneto-viscoelastic theories \cite{saber2026thermo,anand2026coupledtheories}.

\noindent From a computational perspective, robust finite-element implementations, well-defined benchmark problems and reliable parameter-identification strategies are still required to make these models more predictive for engineering design \cite{vogel1,vogel4,pelteret1,haldarKieferMenzel2016,kadapa2022,lijmps2025}. This is especially important for applications involving thin structures, localised magnetic fields, surrounding free-space effects, contact, instability, soft robotic motion and programmed magnetic architectures. Future progress will likely require closer integration between continuum mechanics, micromechanics, experimental characterisation, numerical implementation and data-driven identification. Such integration will support the development of next-generation constitutive models with improved predictive capability, broader applicability and stronger relevance to sensors, actuators, adaptive structures, biomedical devices, vibration-control systems and soft robotic technologies.

\section*{Acknowledgments:} 

\noindent  Authors acknowledge  the support from the Engineering and Physical Sciences Research Council (EPSRC) under the grant (EP/Z535710/1). M Hossain acknowledges the Royal Society-NSFC International Exchange Grant (IEC/NSFC/211316).
\bibliographystyle{elsarticle-num}
\bibliography{Ref}

\end{document}